\documentclass[%
 aip,
 prb,
 amsmath,amssymb,superscriptaddress, 
 reprint,%
]{revtex4-1}

\usepackage{graphicx}
\usepackage{dcolumn}
\usepackage{bm}

\usepackage[utf8]{inputenc}
\usepackage[T1]{fontenc}
\usepackage{etoolbox}

\makeatletter
\def\@email#1#2{%
 \endgroup
 \patchcmd{\titleblock@produce}
  {\frontmatter@RRAPformat}
  {\frontmatter@RRAPformat{\produce@RRAP{*#1\href{mailto:#2}{#2}}}\frontmatter@RRAPformat}
  {}{}
}%
\makeatother
\usepackage{xcolor}
\usepackage{soul}

\begin{document}


\title{Nonequilibrium free energy during polymer chain growth}
\author{Michael Bley}
 \email{michael.bley@physik.uni-freiburg.de}
\affiliation{Applied Theoretical Physics – Computational Physics, Physikalisches Institut, Albert-Ludwigs-Universität Freiburg, Hermann-Herder Strasse 3, D-79104 Freiburg,Germany}%
\author{Joachim Dzubiella}%
 \email{joachim.dzubiella@physik.uni-freiburg.de}
 \affiliation{Applied Theoretical Physics – Computational Physics, Physikalisches Institut, Albert-Ludwigs-Universität Freiburg, Hermann-Herder Strasse 3, D-79104 Freiburg,Germany}
 \affiliation{Cluster of Excellence livMats@FIT - Freiburg Center for Interactive Materials and Bioinspired Technologies, Albert-Ludwigs-Universität Freiburg, Georges-Köhler-Allee 105, D-79110 Freiburg, Germany}%

\date{\today}

\begin{abstract}
During fast diffusion-influenced polymerization, nonequilibrium behavior of the polymer chains and the surrounding reactive monomers has been reported recently. Based on the laws of thermodynamics, the emerging nonequilibrium structures should be characterisable by some 'extra free energy' (excess over the equilibrium Helmholtz free energy). Here, we study the nonequilibrium thermodynamics of chain-growth polymerization of ideal chains in a dispersion of free reactive monomers, using off-lattice, reactive Brownian Dynamics (R-BD) computer simulations in conjunction with approximative statistical mechanics and relative entropy (Gibbs-Shannon and Kullback-Leibler) concepts. In case of fast growing polymers, we indeed report increased nonequilibrium free energies $\Delta F_{\rm neq}$ of several $k_{\mathrm{B}}T$ compared to equilibrium and near-equilibrium, slowly growing chains. Interestingly, $\Delta F_{\rm neq}$ is a non-monotonic function of the degree of polymerization and thus also of time. Our decomposition of the thermodynamic contributions shows that the initial dominant extra free energy is stored in the nonequilibrium inhomogeneous density profiles of the free monomer gas (showing density depletion and wakes) in the vicinity of the active center at the propagating polymer end. At later stages of the polymerization process, we report significant extra contributions stored in the nonequilibrium polymer conformations. Finally, our study implies a nontrivial relaxation kinetics and 'restoring' of the extra free energy during the equilibration process after polymerization.
\end{abstract}

\maketitle

\section{\label{sec:level1} Introduction}

Recently, we studied the nonequilibrium chain growth of single polymers by particle-resolved, reactive Brownian Dynamics (R-BD) simulations.~\citep{Bley2021} We demonstrated that fast nonequilibrium processing pathways lead to remarkably extended structures showing enhanced scaling exponents $\nu$ for the end-to-end distances, $R \propto N^{\nu}$, as a function of the degree of polymerization, $N$.  In particular, we found good-solvent scaling, $\nu \approx 3/5$, although we polymerized in ideal or $\theta$-conditions.  For slow polymerization, we observed the expected $\nu \approx 1/2$ random walk behavior for ideal or $\theta$-chains in equilibrium.~\cite{Rubinstein2003}  Different to the stretching of Rouse chains by an external force,~\cite{Liphardt2002,Seifert2005} the nonequilibrium extended, and thus 'stressed' polymer conformations resulted from spatio-temporal correlations of both the chain and the free reactive monomers in fast, diffusion-influenced reactions for which monomers and chain segments had not enough time to relax during processing. 

In this work, we wish to address the possible thermodynamic consequences of these intriguing nonequilibrium structures. As is well known, the conformational entropy controls various aspects of polymer chains and arises from the amount of different pathways a linear polymer chain can sample for a given end-to-end vector $\bm{R}$.~\cite{DeGennes1979, Rubinstein2003} External and chemical forces can cause significant modifications of the typical Gaussian probability distributions of end-to-end distances $P(R)$ with $R=|\bm R|$, and more extended chains with fixed $N$ decrease the number of accessible realizations and thus lead to a decrease in conformational entropy with $\Delta S_{\mathrm{conf}} < 0$. Hence, out-of-equilibrium chains temporarily store a significant amount of free energy as long as they remain in an entropically unfavorable state. Upon relaxation back to their corresponding equilibrium state, free energy is released from the chain(s) through work or dissipation on different time scales and at different magnitudes.~\cite{Rouse1953, Rubinstein2003, Strobl2007, Alexander-Katz2009, Einert2011, Latinwo2014, Reiter2020, Ramezani2020, Chandran2019} 

In fact, it has been shown experimentally,~\cite{Ramezani2020} for example, that polymer melts (Polystyrene filaments) are capable in lifting macroscopic loads under certain preparation conditions. It is thus envisioned that elaborate synthesis and processing pathways can govern the magnitude of the 'extra' free energy (i.e., excess over equilibrium) of non-equilibrated polymer chains.~\cite{Chandran2019} In terms of information theory such processes encode so-called 'memory' in polymers.~\cite{Reiter2020} This memorized information of a chain is the stored information of in-equilibrium unlikely or inaccessible conformations, and can be retrieved through suited stimuli. These preparation-induced meta-stable states can exhibit long lifetimes and require high temperatures for sufficiently fast relaxation to return to equilibrium states. 

In general, growing attention has been drawn to investigate nonequilibrium polymer properties, how to conserve and control them,  and their consequences on material design. In particular, the possibility emerges to potentially harvest the stored, extra free energy, and related stresses and memory for the design of highly responsive, interactive, or even adaptive materials.~\cite{Stuart2010,Thomas2011, Liu2018, Walther2019} The anticipated wide range of new structural, dynamical, and mechanical properties arises from the nonequilibrium competition between the time scales of (reaction) synthesis, processing, polymer relaxation, and observation and function.~\cite{DeGennes1982a, Latinwo2014, Chandran2019, Reiter2020, Guerin2012, Katkar2018, Chubak2020, Bley2021} However, it is difficult from a theoretical point of view how to define and calculate a nonequilibrium free energy of a growing polymer chain.  Apart from the general challenges to define notions like entropy and heat in nonequilibrium,~\cite{DeGroot1962, Prigogine1978, Seifert2005b, Seifert2012, Evans1993, Jarzynski1997, Jarzynski1997b, Crooks1999} numerical, particle-resolved studies are hampered by the fact that one has to gather time-dependent averages of a complex system with spatiotemporal inhomogeneities. 

In this work, we attempt to address this problem and use the model framework of our previous study~\citep{Bley2021} to investigate the thermodynamics of polymer topologies in nonequilibrium. Our model is appropriately minimalistic (e.g., ideal gas, ideal chains, harmonic bonds) that a nonequilibrium analysis stays relatively transparent with well known equilibrium limits.  The formation of (covalent) bonds during the polymerization process can be characterized as an irreversible thermodynamic process and the growing chain can be considered an entropy sink (energy is put in or generated, while entropy decreases), which follows the second law of thermodynamics.~\cite{DeGroot1962, Prigogine1978, Seifert2012} To calculate the entropy change during the polymerization we then use the most accepted concepts of the definition of a nonequilibrium entropy based on the works of Gibbs~\cite{Gibbs1902} and Shannon~\cite{Shannon1948}, leading to the introduction of a concept of 'relative entropy' (or Kullback-Leibler divergence~\cite{Kullback1951}) for the comparison of equilibrium to nonequilibrium distributions.~\cite{Altaner2017} 

In fact, in polymer physics, Wall derived already 80 years ago the 'Wall equation' for the change in entropy under certain nonequilibrium conditions,~\cite{Wall1942} revisited later from various perspectives~\cite{Dayantis1994, Qian2001} and consistent with the Gibbs-Shannon and Kullback-Leibler forms. It was recently applied, for example, for calculating the nonequilibrium entropy of a flowing polymer melt via atomistic simulations,~\cite{Edwards2021} or the kinetics and thermodynamics of reversible polymerization in closed systems using a master equation approach.~\cite{Lahiri2015} Moreover, direct experimental measurements of a colloidal particle kept in weak nonequilibrium with a virtual double-well potential created by a feedback trap recently demonstrated the utility of the Gibbs-Shannon and Kullback-Leibler concepts for another stochastic molecular system.~\cite{Gavrilov2017}

Here, we employ these concepts to compare nonequilibrium polymer conformations and the heterogeneous distributions of the surrounding free monomer environment with the corresponding equilibrium reference states to calculate the relative energetic and entropic differences, culminating in the nonequilibrium extra free energy $\Delta F_{\mathrm{\rm neq}}(N)$ as a function of the time-dependent degree of polymerization $N(t)$.  For a full understanding of the system we also consider the equilibrium entropic and energetic cost of transferring free monomers into a polymer chain with respect to the formed (harmonic) bonds. Our final results show significant contributions from both nonequilibrium polymer conformations as well as free monomer distributions to the total nonequilibrium transfer free energy (work) in the system, and suggests a complex time dependence of this extra free energy as well of its relaxation kinetics back to equilibrium after the termination of the polymerization.

\section{Model and simulations}

\begin{figure}
\includegraphics[width = 8.5cm]{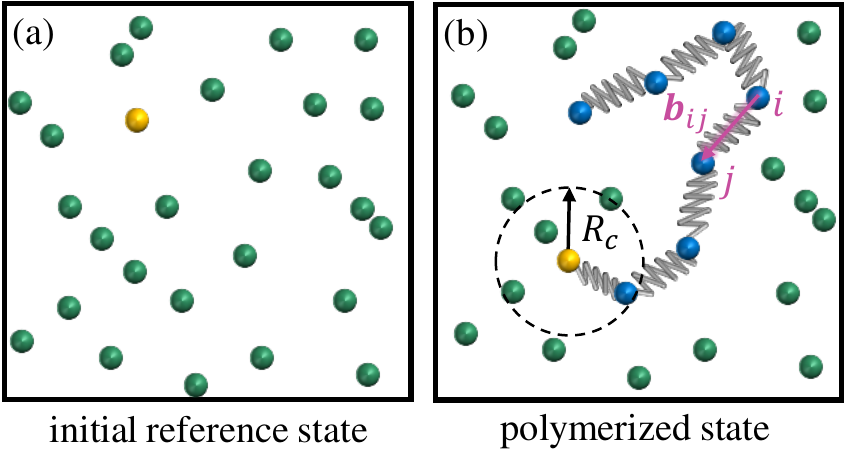}
\caption{\label{Fig1:fig}(a) Initial reference state for polymerization: Free (non-bonded) monomers (green) constitute a reactive ideal gas of initial density $\rho_0 = (M_{0})/V$. The yellow bead is the polymerization seed, i.e., the initial active center (AC) $N(0)=1$ for polymerization towards $N>1$. (b) Intermediate polymerized state:  A few free monomers of (a) became part of the polymer of $N(t)$ monomers with an AC (yellow) of a reactive radius $R_c$ and $N-1$ reacted chain beads (blue) connected through harmonic bonds (gray) with bond vectors $\bm{b}_{ij}$ (pink).}
\end{figure}

Our chain polymerization study is based on reactive Brownian dynamics (R-BD) computer simulations as described in detail previously.~\cite{Bley2021} All simulations have been performed with the LAMMPS package \cite{Plimpton1995} using a standard overdamped BD algorithm for the translational dynamics. In our model, we consider $M(t) = M_{0} - N(t) - 1$  free reactive monomers and $N(t)$ polymer monomers in a fixed volume $V$ at time $t$. $N(t)$ is the time-dependent degree of polymerization. The initial amount of free monomers at time $t=0$ is $M(0)=M_{0} = 1000$ and in addition we consider one polymer seed $N(0)=N_{0} = 1$, cf. Fig.~\ref{Fig1:fig}(a). All monomers (free and polymer) are not interacting, so that we consider the free monomers as an ideal gas, and the polymers as ideal chains in our study.   

The free monomers can diffuse and react with one reactive polymer end -- called in the following the active center (AC) -- to form and grow a single linear chain of $N(t) > 1$ polymer monomers at time $t$, cf. Fig.~\ref{Fig1:fig}(b). The diffusion coefficient for all monomers (free and bound within the chain) is set to $D_{0} = \sigma^{2}/\tau_{\mathrm{B}}$. The size $\sigma = 1.0$ defines our unit for length. The Brownian time $\tau_{\mathrm{B}}$ fixes our unit of time. We employ an integration time step of $\Delta t = 10^{-5}\tau_{\mathrm{B}}$. We simulate at fixed temperature $T$, and our energy unit is the thermal energy $k_{\mathrm{B}}T$, with $k_{\mathrm{B}}$ being the Boltzmann constant. Our statistical ensemble is thus canonical with  constant $V$, $T$, and the total amount of all (free and polymer)  monomers $M_0+1$.

The bonds between two chain monomers $i$ and $j$ in the polymer (see again Fig.~\ref{Fig1:fig}(b)) are described through a harmonic bond potential of the form 
\begin{equation}
u_{ij}(b_{ij}) = \frac{K_{\mathrm{bond}}}{2}(b_{ij} -b_{0})^{2},
\label{eq:bond}
\end{equation}
where $b_{ij}=|\bm b_{ij}|$ is the magnitude of the bond vector, $K_{\mathrm{bond}} = 20 \varepsilon\sigma^{-2}$ is the spring constant, and $b_{0} = 1.0 \sigma$ is the zero-temperature bond length. We fix the energy to $\varepsilon=1k_{\mathrm{B}}T$. No other bonded and non-bonded interactions are considered here, as in the ideal-chain Rouse model.~\cite{Rouse1953,Doi1986} Note, however, that there are two important differences to the Rouse model: i) we interpret our bonded potentials as energetic, not entropic, thus they store potential energy. ii) we employ a non-vanishing, zero-temperature bond extension $b_{0}$ with a relatively stiff spring constant. Hence, our chain model is actually closer to a freely jointed chain (FJC) model~\cite{Rubinstein2003, Strobl2007} (with segment length $b_{0}$) than to the Rouse chain.~\cite{Rouse1953,Doi1986} 

The polymer grows with a prescribed reaction probability if a free gas monomer is close to  the AC, depicted as yellow bead in Fig.~\ref{Fig1:fig}(b). For simulating the chain growth for the reaction $N \rightarrow N + 1$, the algorithm checks every $(\lambda\Delta t)^{-1}$ integration time steps with a probability $p_{\mathrm{react}} = 1.0$ if a reaction is possible. Here, $\lambda$ has the unit per time (i.e., is a rate) and is the reaction propensity following the Doi scheme. \cite{Doi1975a, Erban2009} The parameter $\lambda$ thus controls the speed of the polymerization.  A cut-off protocol method \cite{DeBuyl2015} as implemented in the LAMMPS software package \cite{Plimpton1995} checks if a bond formation between the closest free monomer found within a spherical reactive volume of radius $R_c = \sqrt[6]{2} \sigma \approx 1.122 \sigma$ around the AC is possible. Once the a new bond has been formed, the newly added monomer becomes the AC, and the previous AC is deactivated for any further reaction.  

We consider in our study two reaction speeds: 'slow' and 'fast', as characterized in detail previously.~\cite{Bley2021} The fast reaction has a reaction propensity $\lambda \Delta t = 10^{-3}$ ($\lambda \tau_{\mathrm{B}} = 10^2$), while for slow simulations $\lambda \Delta t = 10^{-5}$ ($\lambda \tau_{\mathrm{B}} = 1$).  For fast growth simulations, more than 4000 different trajectories of a total simulation length of $5 \tau_{\mathrm{B}}$ have been collected and analyzed, whereas for the slow growth 350 simulations of length $250 \tau_{\mathrm{B}}$ were considered.   

We fix the volume of the periodically repeated cubic simulation box to $V = (15\sigma)^3$. In all our polymerization simulations the initial density of free monomers $\rho_{0} = M_{0}/V \simeq 0.297 \sigma^{-3}$ with $M_{0} = 1000$ is used.  We also conduct equilibrium simulations of chains with fixed $N = 10$, $20$, $50$, $100$, $150$, $250$, $500$ and $1000$ and the corresponding monomer number $M = M_0 - N -1$ for $50$ up to $5000~\tau_{\mathrm{B}}$ to compare one-to-one to the nonequilibrium situation characterized by $M(t)$ and $N(t)$. 

\section{Nonequilibrium polymerization thermodynamics}

Our thermodynamic analysis of the changes of the free energy in the system during polymerization is based on the following considerations. At first, we consider an unpolymerized equilibrium reference state (Fig.~\ref{Fig1:fig}(a)), which has an equilibrium (Helmholtz) free energy $F_0$. We define the (total) nonequilibrium free energy difference to a polymerized state with $N(t) > 1$ chain monomers and free energy $F(N)$, as in Fig.~\ref{Fig1:fig}(b), as
\begin{equation}
\begin{split}
\Delta F (N) = F(N)-F_0 = \Delta F_{\rm eq}(N) + \Delta F_{\rm neq}, 
\end{split}
\label{Eq1:eq}
\end{equation}
where $\Delta F_{\rm eq}(N) = F_{\rm eq}(N) - F_0$ is the equilibrium work of forming the polymer state.  Hence, we expect $\Delta F(N) \rightarrow \Delta F_{\rm eq}(N) $ for very slow, quasi-stationary polymerization. The true nonequilibrium contributions are in $\Delta F_{\rm neq} = \Delta F - \Delta F_{\rm eq}$. Further, we assume we that we can split $\Delta F_{\rm neq}$ into a 'polymer' part and a 'monomer gas' part, via
\begin{equation}
\Delta F_{\rm neq} =  \Delta F_{\rm pol} + \Delta F_{\rm gas}, 
\label{eq:extra}
\end{equation}
as discussed in detail in the forthcoming sections. The nonequilibrium free energy $\Delta F_{\rm neq}$ we interpret as the 'extra' or 'stored' free energy in nonequilibrium, i.e., the free energy in excess over equilibrium. As motivated in the introduction, the contributions in Eq.~(\ref{eq:extra}) we will calculate in terms of Gibbs-Shannon and Kullback-Leibler concepts from nonequilibrium distributions. 

According to the additive assumptions of~Eq.(\ref{Eq1:eq}), the explicit consideration and evaluation of $\Delta F_{\rm eq}$ is in principle not necessary in our work, since $\Delta F_{\rm neq}$ will be evaluated independently as the deviation from the (in simulation calculated) equilibrium distributions. However, we inspect $\Delta F_{\rm eq}$ also in detail in to understand the thermodynamics of the modelled process in a more comprehensive fashion.  Note that expression~(\ref{Eq1:eq}) can in principle also be considered as a function of time $t$ instead of $N$, because $N(t)$ can be directly accessed from simulations (or from experiments) or modelled through a reaction rate law, if available.~\cite{Bley2021}

\subsection{Equilibrium transfer free energy: $\Delta F_{\rm eq}$}

With $\Delta F_{\rm eq}$ we denote the equilibrium free energy difference of transferring a free monomer from the bulk of accessible volume $V$ (and density $\rho(t)=M(t)/V$) into the polymer chain of length $N$.  We approximate $\Delta F_{\rm eq}$ by the following expression derived from equilibrium partition sums (see section S1 in the supplementary material),
\begin{equation}
\Delta F_{\rm eq}(N) \simeq - k_{\mathrm{B}}T \left[ (N-1) \ln \left( \frac{v_{0}}{V} \right) + \ln \frac{M_{0}!}{N!M!}  \right] .
\label{eq:DF1}
\end{equation}
Using the Stirling approximation $\ln(n!)\simeq n\ln(n)-n$, and assuming $N\gg1$ and $M_0\gg N$, this can be written in a more familiar way, as: 
\begin{equation}
\Delta F_{\rm eq} \simeq -Nk_{\mathrm{B}}T\left [\ln\left(\rho\frac{v_0}{N}\right) - 1\right] . 
\label{eq:DF1S}
\end{equation}
Thus, it describes essentially the free energy cost of confining $N$ free ideal particles of bulk density $\rho$ within the polymer chain with an effective density $N/v_0$.  The nature and definition of the volume $v_0$ will be motivated and explained in the following. In our evaluation we use the form (\ref{eq:DF1}) and not the more approximate expansion (\ref{eq:DF1S}), because we are not always strictly in the Stirling approximation regime. 

In the derivation of $v_0$, we follow earlier work on the partition sum of a slightly extensible FJC model, i.e., a bonded polymer like in our simulation model with relatively stiff harmonic bonds.~\cite{Winkler,Colombo,Fiasco}  The $N-1$ bond degree of freedoms of the polymer chain decouple and per bond one can define the internal partition sum $v_{0}$ (in units of volume), according to 
\begin{equation}
v_{0} = 4\pi \int_{0}^{\infty} \mathrm{d}b \, b^{2} \exp \left[ -(b-b_{0})^{2}/(2\sigma_{\mathrm{b}}^{2}) \right],
\label{eq:v0}
\end{equation} 
where $\sigma_{\mathrm{b}}= (2K_{\mathrm{bond}}\beta)^{-1/2}$ is the temperature-dependent standard deviation of the Gaussian probability distribution, and the factor 2 in front of the simulation spring constant reflects the action of the two neighboring potentials. Evaluating Eq.~(\ref{eq:v0}) in the limit of large and stiff bonds ($b\simeq b_0 \gg \sigma_{\mathrm{b}}$), we find the known result~\cite{Winkler,Fiasco} 
\begin{equation}
v_{0} \approx 4\pi b_0^2 \sqrt{2\pi}\sigma_{\mathrm{b}},
\label{eq:v0_explicit}
\end{equation} 
which essentially describes bond vector rotations on a sphere of radius $4\pi b_0^{2}$ as in the FJC model,~\cite{Strobl2007} and, additional small longitudinal vibrations of size $\sqrt{2\pi}\sigma_{\mathrm{b}}$. 

For the estimate of $\Delta F_{\rm eq}$, Eq.~(\ref{eq:DF1}), from simulations, we assume that the stiff spring limit Eq.~({\ref{eq:v0_explicit}) applies, but with values of $b_0$ and $\sigma_{\rm b}$ replaced by average bond lengths $\bar b$ and their standard deviations $\bar \sigma_{\rm b}$ in the simulations. The reason is that it is {\it a priori} unclear if the assumptions leading to Eq.~(\ref{eq:v0_explicit}) are valid in nonequilibrium and the bond values remain the same as predicted by theory.  The resulting bond values are summarized in Table~\ref{Tab1:tab}. As we see, the values hardly depend on the process (equilibrium versus slow versus fast polymerization), demonstrating that our statistical mechanics model is meaningful to estimate the equilibrium free energy $F_{\rm eq}$ contribution during the polymerization process.  The average simulated bond length is indeed close to $b_0$ and around $\bar b \simeq 1.04\sigma$, while the width  is $\sigma_{\mathrm{b}} \simeq 0.16\sigma$,  highly consistently comparing to the stiff spring limit $\sigma_{\mathrm{b}}= (2K_{\mathrm{bond}}\beta)^{-1/2} = (1/\sqrt{40})\sigma\simeq 0.16\sigma$.

\begin{table}
\caption{\label{Tab1:tab}Fitted mean bond lengths $\bar{b}$ and standard deviations $\bar{\sigma}_{\mathrm{b}}$ for selected chain lengths $N$ for equilibrium systems (eq), and slow and fast growth.}
\begin{ruledtabular}
\begin{tabular}{ccccccc}
$N$ & $\bar{b}_{\mathrm{eq}}\sigma^{-1}$ & $\bar{\sigma}_{\mathrm{b,eq}} \sigma^{-1}$ & $\bar{b}_{\mathrm{slow}}\sigma^{-1}$ & $\bar{\sigma}_{\mathrm{b,slow}} \sigma^{-1}$ & $\bar{b}_{\mathrm{fast}}\sigma^{-1}$ & $\bar{\sigma}_{\mathrm{b,fast}} \sigma^{-1}$ \\
\hline 
20 & 1.04 & 0.16 & 1.04 & 0.15 & 1.03 & 0.16 \\ 
50 & 1.04 & 0.16 & 1.04 & 0.16 & 1.04 & 0.16 \\ 
100 & 1.04 & 0.16 & 1.04 & 0.16 & 1.05 & 0.16 \\ 
150 & 1.04 & 0.16 & 1.05 & 0.16 & 1.05 & 0.16 \\ 
200 & 1.04 & 0.16 & 1.04 & 0.16 & 1.05 & 0.16 \\ 
250 & 1.04 & 0.16 & 1.04 & 0.15 & 1.05 & 0.16 \\ 
\end{tabular}
\end{ruledtabular}
\end{table}

\begin{figure*}[ht!]
\includegraphics[width = 17cm]{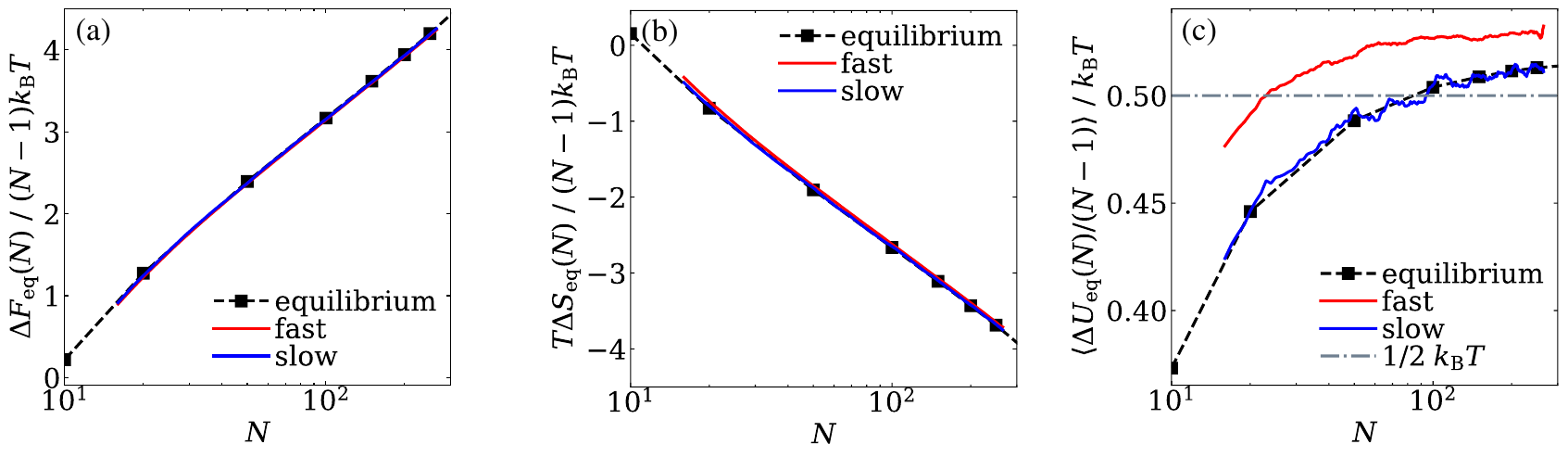}
\caption{\label{Fig2:fig} (a) Equilibrium transfer free energy $\Delta F_{\rm eq}(N)$, (b) entropic  contribution $T\Delta S_{\rm eq}(N)$, and (c) internal energy $\Delta U_{\rm eq}(N)$ per bond as a function of the chain length $N$. The black symbols represent results from non-growing, equilibrium chains, and the fast ($\lambda \Delta t = 10^{-3}$) and slow reactions ($\lambda \Delta t = 10^{-5}$) are depicted as red and blue lines. The gray dash-dotted line in (c) represents the $(N-1) k_{\mathrm{B}}T/2$ contribution, as estimated from the statistical mechanics model, Eq.~(\ref{eq:U}).}
\end{figure*}

From the free energy, Eq.~(\ref{eq:DF1}), the equilibrium entropy change follows from $\Delta S_{\rm eq} = - \partial \Delta F_{\rm eq}/\partial T$, and we obtain 
\begin{equation}
\Delta S_{\rm eq}(N) = - \Delta F_{\rm eq}(N)/T + (N-1)k_{\mathrm{B}}/2.
\end{equation}
The difference in internal energy $\Delta U_{\rm eq}(N) = \Delta F_{\rm eq}(N) + T \Delta S_{\rm eq}(N)$ is then given by 
\begin{equation}
\Delta U_{\rm eq}(N) = (N-1)k_{\mathrm{B}}T/2.
\label{eq:U}
\end{equation} 
This result for the energy is understandable in terms of $N-1$ independent harmonic oscillators which act as linearly vibrating diatoms.~\cite{McQuarrie1976} The vibrational degree of freedom partitions $k_{\mathrm{B}}T/2$ into the kinetic energy which cancels in the difference $\Delta F_{\rm eq}$ with one translational degree of freedom of the ideal monomer gas. Equally, one $k_{\mathrm{B}}T/2$ goes in the potential energy which remains as net effect. For evaluating  $\Delta U(N)$ from the simulations, we simply average the internal bond energy for a fixed $N$ over nonequilibrium trajectories. In that sense, $\Delta U(N)$ is an average in nonequilibrium~\cite{Altaner2017} in our polymerization simulations and should be denoted as $\Delta U_{\rm neq}(N)$. However,  we will see that is has negligible nonequilibrium contributions and we can safely interpret it as $\Delta U_{\rm eq}(N)$.

Fig.~\ref{Fig2:fig}(a) shows the resulting curves from the simulation averages for $\Delta F_{\rm eq}(N)$ per bond (i.e., divided by $N-1$). Entropic and energetic contributions are presented in panels (b) and (c), respectively.  We see that the free energy cost $\Delta F_{\rm eq}(N)$ per formed bond is on the order of several $k_{\mathrm{B}}T$, mostly contributed ($\gtrsim 90\%$) from entropy as seen from  panel (b).   Hence, the formation of a new harmonic bond leads to a significant loss of entropy  for the added particles to the polymer chain from the ideal gas state, since the particle's accessible volume and thus configurational freedom decreases drastically from $V$ to $v_{0}$.  This penalty per bond increases with $N$ (or, with time) because the density of the free monomer gas decreases with polymerization in our canonical ensemble. The contribution of the internal energy $\Delta U_{\rm eq}$ (Fig. \ref{Fig2:fig}c) is close to $k_{\mathrm{B}}T/2$ in very good agreement with our statistical predictions above. It is thus only a minor contribution to the free energy change. (We observe a slight but systematic increase of $\Delta U_{\rm eq}(N)$ per bond in the simulations, possibly because for shorter chains small finite length effects play a role in the equi-partitioning among the vibrational modes. Note also that for larger $N\gtrsim 100$ the equilibrium simulation saturates at values about 5\% larger than $k_BT/2$. This is very probably due to the approximations made when going from Eq.~(\ref{eq:v0}) to Eq.~(\ref{eq:v0_explicit}) and the $k_BT/2$ limit is not exact.)

Importantly, the free energy cost estimated here for all processes is essentially the same, in particular the differences between fast and slow polymerization reactions with respect to equilibrium are negligibly small on the $k_BT$ scale. Hence, nonequilibrium effects are not significant in our evaluation of $\Delta F_{\rm eq}(N)$ which validates our treatment to estimate a true equilibrium contribution of the free energy of polymerization. We will in the following better understand where the nonequilibrium contributions in the polymer are, when we discuss $F_{\rm pol}$, i.e., the nonequilibrium free energy stored in the polymer conformations. 

Note that our model reaction would be characterized as 'endogenic' and 'endothermic',~\cite{Atkins2010} because $\Delta F_{\rm eq}>0$ and $\Delta U_{\rm eq}>0$, respectively, i.e., the reaction is not spontaneous and leading to higher system energies. In our model it could be simply made 'exo-' (spontaneous, and producing energy) by shifting the bonding energy by a constant to very favorable negative values. The following analysis on the nonequilibrium contribution $\Delta F_{\rm neq}$ would not be influenced by such a global energetic shift. 

An interesting final remark in this section concerns the statistical $1/N!$ 'Gibbs factor' in  Eq.~(\ref{eq:DF1}). We include the latter (as in the conventional Boltzmann statistics of classical gases~\cite{McQuarrie1976})  because we assume in our system that the $N$ polymer beads are indistinguishable. Hence, in our interpretation, an exchange of two beads does not yield a different microstate we want to consider. In other words, upon particle exchange in the polymer the system information (relative entropy) stays the same in our interpretation.  The analysis framework and main conclusions of our study on the behavior of $F_{\rm neq}$ do not change by neglecting the Gibbs factor. (The consequence would be the absence of the $N$-factor in the logarithm in Eq.~(\ref{eq:DF1S}) and thus a systematic change and shift of the $N$-dependence of $\Delta F_{\rm eq}$).    

\subsection{Polymer conformational entropy in nonequilibrium: $\Delta F_{\rm pol}$}

Per construction and our interpretation, the term $\Delta F_{\rm eq}$ solely contains the changes in equilibrium free energy originating from the internal bond confinement and longitudinal distributions while conserving full rotational freedom per segment. Nonequilibrium effects, however, originate from rotational constraints of the bonds, which do appear in nonequilibrium. These rotational correlations lead to deviations from ideal chain statistics, as for the FJC model, in particular, leading to chain elongation and the previously reported remarkable self-avoiding walk  scaling for ideal chains.~\cite{Bley2021}  

The enhanced size scaling in nonequilibrium polymerization leads to shifted probability distributions $P(R(N))$ for the end-to-end distances $R(N)$. Examples for emerging probability distributions $P(R(N))$ for equilibrium and slow and fast reaction conditions for $N = 50$ and $N = 150$ are presented in Fig.~\ref{Fig3:fig}. The distributions for fast reactions are indeed substantially shifted to more extended nonequilibrium conformations for both chain lengths. 
     
\begin{figure}[ht!]
\includegraphics[width = 8.5cm]{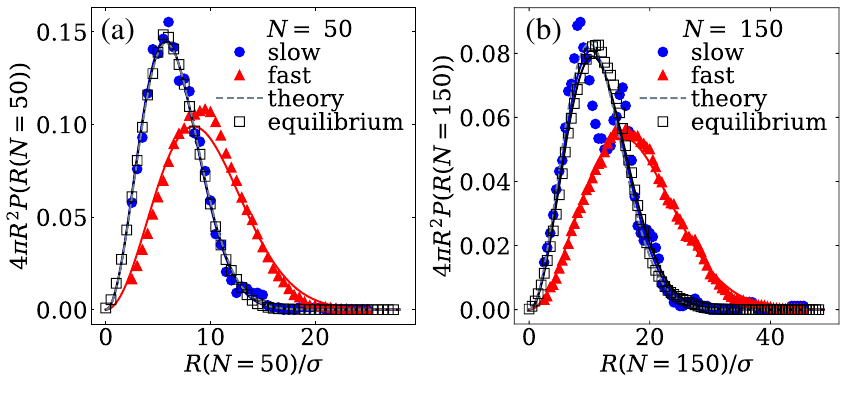}
\caption{\label{Fig3:fig} (a) and (b) Normalized distributions of end-to-end distances $P(R(N))$ from simulation (symbols) and Gaussian fits (solid lines). The equilibrium case is compared with the theory (gray dashed line) for ideal chains~\cite{Rubinstein2003, Strobl2007} at chain lengths $N = 50$ and $150$, respectively. Note that we plot the distribution with the appropriate spherical weight, $4\pi r^2 P(R)$, so that the quality of the fits of the tails of the distributions is visible.} 
\end{figure}

The distributions $P(R)$ can be used for calculating the relative entropy $\Delta S_{\rm pol}$ (also, Kullback-Leibler divergence) \cite{Kullback1951, Altaner2017} for accessing the $\Delta F_{\rm pol}$-term of Eq.~(\ref{Eq1:eq}). While we employ the end-to-end distance,  we believe a more accurate treatment would need to address all individual microstates, i.e., the (translational and rotational) distributions of every individual bond vector and evaluate them using the Kullback-Leibler relative-entropy form.\cite{Qian2001} This is, however, statistically very challenging. We chose therefore the end-to-end distance following the arguments of Dayantis.\cite{Dayantis1994} He showed that as long as the bond microstates are behaving homogeneous, i.e., have all the same distributions for a given end-to-end distance, then our Eq.(10) should be accurate. However, our chain has some spatial anisotropy because it is stretched more at the growing front than at the relaxing other end. In other words, for a fixed end-to-end distance the microstates are not equally probably but biased by the slightly stretched growing end. Thus, we expect deviations in a more accurate treatment, and likely the real contributions are somewhat larger than currently calculated.

Given these considerations, the relative entropy for the polymer conformations in our work is defined as 
\begin{equation}
\Delta S_{\rm pol} = -4\pi k_{\mathrm{B}} \int_0^\infty  \mathrm{d}R\,R^2 \;P(R(N)) \ln \left[\frac{P(R(N))}{P_{0}(R(N))} \right],
\label{eq:DF2}
\end{equation}
and we use the definition of a nonequilibrium free energy~\cite{Qian2001,Altaner2017}
\begin{equation}
\Delta F_{\rm pol}(N) = -T \Delta S_{\rm pol} 
\label{eq:DF2}
\end{equation}
which is the (purely entropic) free energy difference associated with the nonequilibrium deviation from the equilibrium distribution.~\cite{Dayantis1994} The latter we denote as $P_{0}(R)$.~\cite{Rubinstein2003, Strobl2007} In other words, Eq.~(\ref{eq:DF2}) vanishes in equilibrium, that is, it accounts only for the deviations of information  from equilibrium.  We obtain continuous functions for $P(R(N))$ from fit functions of the simulated histograms (see S2 in the supporting information for the functions and fit parameters). Fig.~\ref{Fig3:fig} demonstrates that Gaussian fits also describe well the nonequilibrium distributions, also accounting for the tails of $4\pi R^2 P(R)$. 

Fig.~\ref{Fig4:fig}(a) presents the resulting  $\Delta F_{\rm pol}(N)$ (filled symbols) for the two reaction speeds. As expected, the nonequilibrium free energy contribution to Eq.~(\ref{Eq1:eq}) is negligibly small for slow reactions, as provided already by the agreement of the $P(R)$ for equilibrium and slow reactions (Fig.~\ref{Fig3:fig}). Fast reactions, however, lead to an almost linear dependency $\Delta F_{\mathrm{pol}}(N) \propto N$ with $\Delta F_{\mathrm{pol}}(N) \approx 1.0~k_{\mathrm{B}}T$ for long chains ($N \approx 250$). This indicates that the more extended chain stores some extra free energy, which can be released when the chain relaxes to equilibrium. A scaling analysis  put forward in our previous paper~\cite{Bley2021} suggested that the relative lengths of polymer segments that can relax during polymerization is decreasing with polymerization time, scaling with $\propto \sqrt{1/N}$. Hence, we expect $\Delta F_{\rm pol}(N)$ to continuously grow with chain length.  

Figure~\ref{Fig4:fig}(b) now plots $\Delta F_{\rm pol}(N)(t)$ as a function of time for fast reactions and shows that this free energy contribution increases in time.  Hence, nonequilibrium polymer chain conformations can store a significant amount of free energy, increasing with polymerization time. However, it is not very large in our model system within the studied observation time, just on the order of one $k_{\mathrm{B}}T$. 

Coming back briefly to the discussion whether the end-to-end distance is a good coordinate for our evaluation. We can give some  complementary, approximative arguments why our current treatment is at least reasonably accurate but probably a lower limit. In our previous paper,~\cite{Bley2021} we showed that the growing bond vectors are directed and instead of rotating full area $4\pi$ (of the unit sphere), only rotate about 80\% of it (as estimated from the accessible azimuthal angles). This leads to an entropic penalty of roughly $-k_BT\ln(0.8)\simeq0.22~k_BT$ per bond. We also demonstrated that around 5 bonds have not relaxed in the early steady state (after one $\tau_B$). This would amount to about $5\times0.22 \simeq 1.1~k_BT$ entropic penalty per chain, growing in time because of less and less relaxation of the chain in the process (see also the arguments later at the end of section III.D). This estimate is larger than (but close to) about~0.6$k_BT$ which we calculate after one $\tau_B$ (where the system is diffusive stationary). Hence, the conformational entropy might be indeed a bit higher than calculated.

\subsection{Free monomer gas translational entropy in nonequilibrium: $\Delta F_{\rm gas}$}

\begin{figure}[ht!]
\includegraphics[width = 8.5cm]{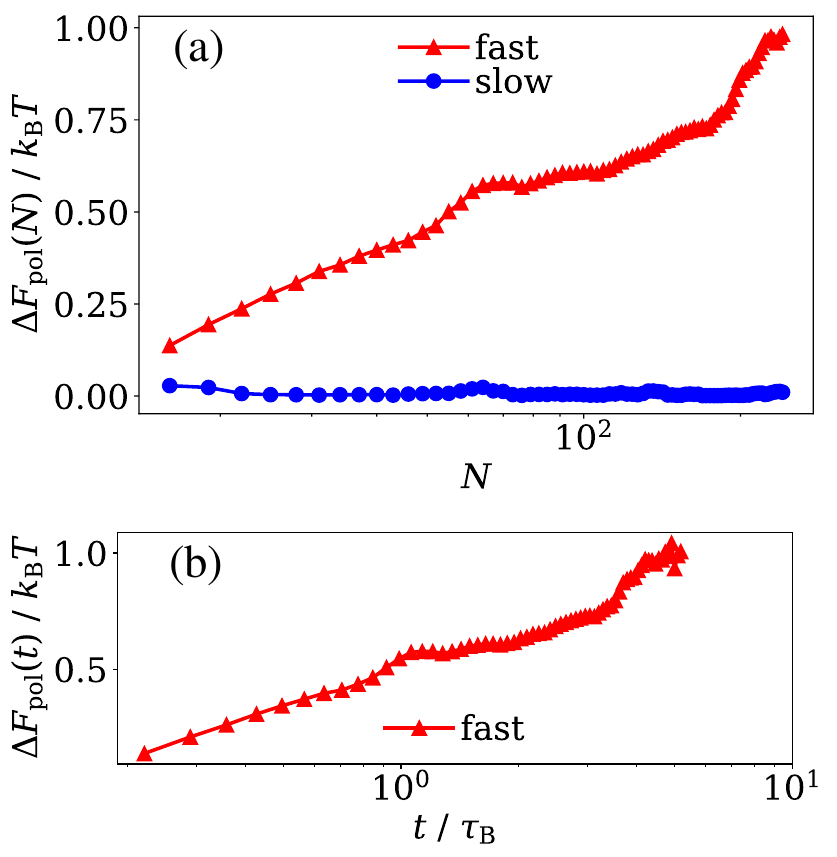}
\caption{\label{Fig4:fig} (a) Nonequilibrium polymer conformational free energy $\Delta F_{\rm pol}(N)$ of a growing chain as function of chain length $N$. (b) $\Delta F_{\rm pol}(t)$ as a function of time $t$ for fast reactions. Evaluated from Eq.~(\ref{eq:DF2}), see details in the text.}
\end{figure}

In addition to the polymer chain, also the free monomer ideal gas exhibits nonequilibrium spatio-temporal correlations.~\cite{Bley2021} Due to the fast reaction at the AC, the free monomer one-body density profile $\rho(r,t)$, where $r$ is the distance to the center-of-mass of the AC, is non-uniform in space and time. A classical, well-known reference is the density depletion ($\rho(r) \propto (1-R_c/r$)) around the spherically-symmetric 'Smoluchowski sink' of radius $R_c$ in {\it steady-state} diffusion-controlled bimolecular reactions.~\cite{Smoluchowski1918} The magnitude of depletion depends on the reaction speed in diffusion-influenced reaction with radiation (Colllins-Kimball) or Doi-type of boundary conditions.~\cite{Dibak2019} 

In order to study and characterize the inhomogeneous monomer distributions, we calculated the non-equilibrium one-body density profiles at fixed $N$, $\rho(r;N)$, as a function of the radial distance $r$ from the AC of the growing chain. In general, the monomer distribution will be heterogeneous in space and a function of absolute space, $\rho(\bm{r} ;N)$, but is difficult to obtain for our anisotropic system -- so we approximate the distribution by considering the average, spherically symmetric profile $\rho(r;N)$ around the AC. The profiles are averaged for fixed $N$ at the time right when the reaction occurs.~\cite{Bley2021} 

Results are presented in Fig.~\ref{Fig5:fig}(a) and (b) for chain lengths $N = 50$ and $150$, respectively. Indeed, faster reaction rates cause a large depleted volume around the AC for $r\lesssim 5\sigma$, qualitatively similar to the classical Smoluchowski density holes with a size set by the reactive sink radius according to $\rho(r) \propto (1-R_c/r$). The nonequilibrium depletion then leaves a wake of holes vanishing in space and time behind the AC.~\cite{Bley2021}} In contrast, for the slow reactions much smaller depletion regions $r\lesssim 2\sigma$ are observable, as we expect from diffusion-influenced Doi-reactions with smaller intrinsic propensities.~\cite{Smoluchowski1918,Dibak2019} For large distances, the nonequilibrium profiles saturate to an almost homogeneous profile at the simulation box boundary, being very close to the homogeneous  equilibrium profile $\rho_{0}(N)$ (independent of $r$) for fixed $N$. 

\begin{figure}[ht!]
\includegraphics[width = 8.5cm]{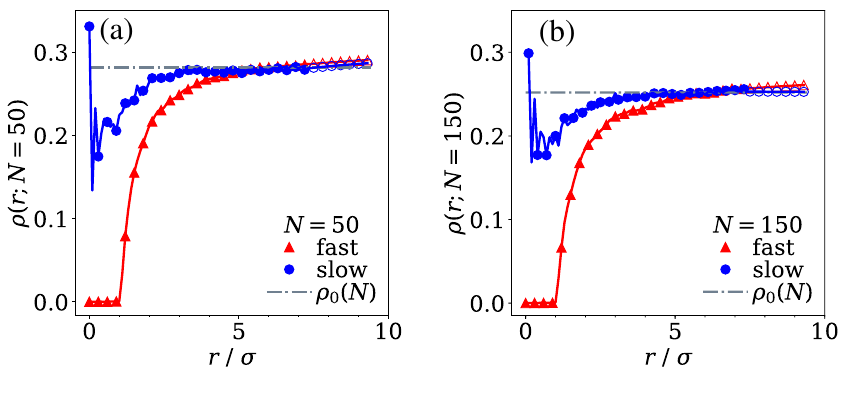}
\caption{\label{Fig5:fig} Nonequilibrium density profiles $\rho(r;N)$ of the free monomer gas at fixed $N$ and radial distance $r$ from the active center $\rho_{r,N}$. Filled symbols are data points from the simulation and empty symbols from extrapolation (see S3 in the supporting information) for chain lengths (a) $N = 50$ and (b) $N=150$. The solid lines represent the interpolation functions between the data points and are used for subsequent integration. The gray dash-dotted lines represent the homogeneous (independent of $r$) equilibrium density $\rho_{0}(N = 50) \approx 0.28~\sigma^{-3}$ and $\rho_{0}(N = 150) \approx 0.22~\sigma^{-3}$.} 
\end{figure}

We now employ again the concept of relative entropy~\cite{Qian2001, Altaner2017} to estimate the free energy cost arising from the density depletion in the non-equilibrium density profiles $\rho(r;N)$, which writes for the density distributions in terms of a nonequilibrium free energy
\begin{equation}
\Delta F_{\rm gas}(N) = -T\Delta S_{\rm gas}=k_{\mathrm{B}}T \int_{0}^{r_{\mathrm{box}}} \mathrm{d} \bm{r} \, \rho(r;N) \ln \left[ \frac{\rho(r;N)}{\rho_{0}(N)} \right],
\label{eq:DF3}
\end{equation} 
where $\rho_{0}(N)$ is the equilibrium reference density for a given remaining monomer concentration. For the equilibrium systems, $\rho(r;N) = \rho_{0}(N)$ for all distances $r$ is valid and thus $\Delta F_{\rm gas}(N) = 0$. 

For numerical evaluation of Eq.~(\ref{eq:DF3}), all non-equilibrium density profiles were extrapolated for distances between $V^{1/3}/2 < r \leq r_{\mathrm{box}}$. This is because the integration of the density profiles refers to a spherical volume which fulfills $M(t) = \int_{0}^{r_{\mathrm{box}}} \mathrm{d} \bm{r} \; \rho(r)$  with $r_{\mathrm{box}} = (3V/(4\pi))^{1/3}$, while the cubic simulation box with periodic boundary conditions allows only calculating up to distances $r= V^{1/3}/2$.  Details on the numerical integration and a sensitivity check of the influence of the nature of the extrapolation can be found in section S3 in the supporting information. 

\begin{figure}[ht!]
\includegraphics[width = 8.5cm]{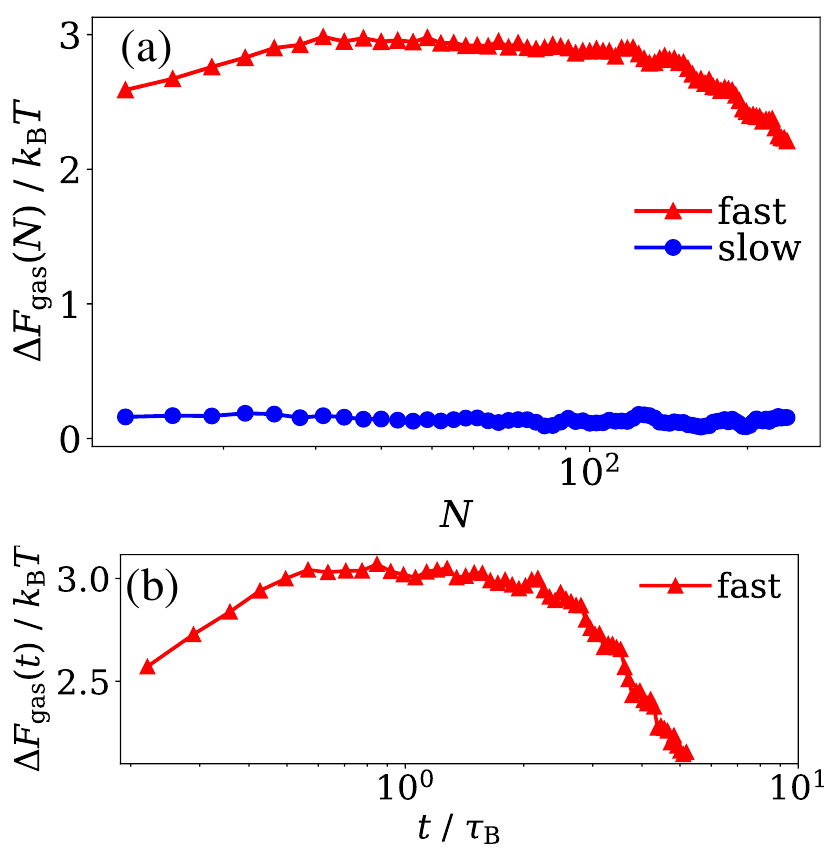}
\caption{\label{Fig6:fig}(a) Nonequilibrium translational free energy $\Delta F_{\rm gas}(N)$ of the free monomer gas around the active center (AC) as a function of chain length $N$. (b) $\Delta F_{\rm gas}(t)$ as a function of time $t$ for fast reactions. Evaluated from Eq.~(\ref{eq:DF3}), see details in the text.}
\end{figure}

The numerical integration yields the $N$-dependent relative entropy for growing chains as plotted in Fig.~\ref{Fig6:fig}(a). We indeed observe substantial nonequilibrium effects, in particular, the magnitude of $\Delta F_{\rm gas}(N)$ is generally large for fast reactions, about $3k_{\mathrm{B}}T$, which is much larger than that for the slow reactions, where only $\Delta F_{\rm gas}(N)\simeq 0.2k_{\mathrm{B}}T$. Hence, the different density profiles $\rho(r;N)$ with respect to the corresponding bulk density $\rho_{0}(N)$ (Fig.~\ref{Fig5:fig}) result in an appreciable difference in stored free energy in nonequilibrium. 

While in the slow polymerization case, $F_{\rm gas}(N)$ is small and roughly constant, the fast case exhibits an interesting non-monotonicity in $N$. It can be better interpreted by inspecting $F_{\rm gas}(t)$ as a function of time $t$ as plotted in~\ref{Fig6:fig}(b): For small times, $t\lesssim \tau_{\mathrm{B}}$, first the depletion hole needs a Brownian (monomer diffusion) time $t \simeq \tau_{\mathrm{B}}$ to build up. Then, there is a maximum of $F_{\rm gas}(t)$ where the system  for several $\tau_{\mathrm{B}}$ is in a steady-state. However, as the reaction progresses and more monomers are added to the growing chain ($t \gtrsim \tau_{\mathrm{B}}$), we report a decrease for $\Delta F_{\rm gas}$, which is linked to the decreasing bulk monomer density and the corresponding diminishing entropic cost of forming the low-density depletion hole. This decrease would not occur in systems with an infinite reservoir of free monomers where the density around the active center remains constant at all times, opposite to our simulations with an initially finite number of to-be-consumed momomers. However, in a typical experiment in a dispersion of many growing polymer chains also the number of monomers per polymer is finite and thus will decrease (be consumed) in time during the reaction process. Hence, such a non-monotonic behavior with a maximum in stored free energy may also occur experimentally. 

\subsection{Nonequilibrium extra free energy}

\begin{figure}[ht!]
\includegraphics[width = 8.5cm]{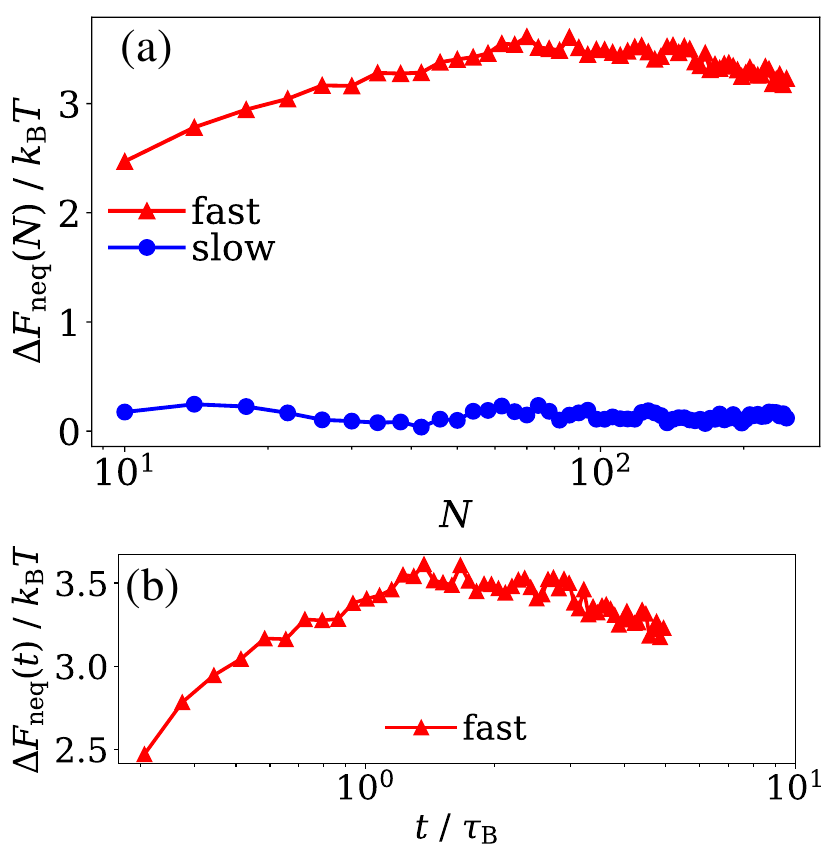}
\caption{\label{Fig7:fig}(a) Nonequilibrium free energy $\Delta F_{\mathrm{neq}}(N)$ of a growing chain as function of chain length $N$. (b) $\Delta F_{\rm neq}(t)$ as a function of time $t$ for fast reactions. Defined as in Eq.~(\ref{eq:extra}).}
\end{figure}

We are now in the position to plot and discuss the extra free energy in nonequilibrium, $\Delta F_{\rm neq} = \Delta F_{\rm pol}+\Delta F_{\rm gas}$. 
$N$-dependent results are plotted in Fig.~\ref{Fig7:fig}(a) for slow and fast reaction speeds. The nonequilibrium contributions to the free energy highly depend on the reaction speed $\lambda \Delta t$. Importantly, for fast reactions up to $3~k_{\mathrm{B}}T$ are stored in the system. Here, for low $N$, $\Delta F_{\mathrm{neq}}(N)$ is dominated by the depletion hole in the ideal monomer gas arising from $\Delta F_{\rm gas}(N)$. Longer chains lead to an increasing contribution from extended chain conformations $\Delta F_{\rm pol}(N)$.

Figure~\ref{Fig7:fig}(b) depicts the corresponding time-dependency of the fast nonequilibrium contribution. After an increase of $\Delta F_{\mathrm{neq}}$ by around $1.0 k_{\mathrm{B}}T$ until $t \approx \tau_{\mathrm{B}}$, the increasing conformational contribution $\Delta F_{\rm pol}$ and the decreasing density contribution $\Delta F_{\rm gas}$ balance out and lead to an almost constant value around $3 k_{\mathrm{B}}T$ per chain. Slow reactions exhibit only a very minor nonequilibrium contribution of the free energy at all times which arises mostly from the small monomer gas depletion around the AC.

We finally discuss briefly the question of what are the possible time scales of relaxation of the polymer back to equilibrium? In other words, how long does it take for the polymer to release the stored free energy, if let freely relaxed after polymerization? We believe there would be a two-step process: 
1) The perturbed nonequilibrium density profile $\rho(r;N)$, which yields $\Delta F_{\rm gas}(N)$ can be linked to a short-term, almost chain-length independent energy release within a $\tau_{\mathrm{B}}$-scale. Hence, the major part of the stored free energy will dissipate relatively quickly back into the system. 
2) The second time scale is set by the polymer relaxation. Given the concept of segmental relaxation of segments $s = N/p$ of mode $p$ of a single chain with the largest segment $s = N$ for $p=1$ as the Rouse time \cite{Rouse1953} with $\tau_{\mathrm{R}} \approx N^{2}b_0^2/(3 \pi^{2} D_{0})$, then $\Delta F_{\rm pol}$ is released gradually back to equilibrium with $\tau_{\mathrm{R}} \propto \tau_{\mathrm{B}} N^2/(3\pi^2)$, which is rather slow for long chains. These considerations may change in very dense, many chain polymer systems, such as melts or glass-like systems.~\cite{Rubinstein2003} 

\section{Concluding Remarks}

In this work, we introduced a framework to analyze data from non-equilibrium BD chain growth simulations to quantify and characterize the nonequilibrium free energy stored in a fast growing ideal polymer chain as a function of chain length $N$ (or time $t$). While we treated the monomer transfer free energy from bulk into the polymer within equilibrium statistical mechanics, the non-equilibrium contributions we expressed using the Gibbs-Shannon / Kullback-Leibler divergence formalisms for relative entropy, applied to the spatial distributions of monomers (bound and free). Here we assumed, arguably, that the end-to-end distance is a good coordinate to characterize conformational distributions. However, this should be checked more carefully in future work. We also assumed that two contributions, one arising from non-equilibrium, extended polymer chain conformations and the other from spatio-temporal correlations of the reaction-perturbed monomer density profiles sum up to the nonequilibrium free energy. 

Within our framework, we demonstrated that a single fast growing chain stores several $k_{\mathrm{B}}T$ of non-equilibrium free energy for all chain lengths $N$. In addition, we observed a non-trivial time-dependency of the nonequilibrium free energy. While the polymer conformational contributions increase continuously in time which can be rationalized by the competition of relaxation and observation time scales,~\cite{Bley2021} those of the free monomers were found to be non-monotonous (with a maximum) due to a short-time built-up of the depletion regions and a long-time decrease of bulk monomer density. Extrapolating to even longer times, we suspect that even a minimum can occur again when the polymer conformational contribution dominates over the free monomer gas contribution. As a consequence, complex kinetics of free energy release related to the time scales of Brownian diffusion and segmental chain dynamics can be envisioned for chain relaxation back to equilibrium after polymerization. We wonder if and how it would be possible in future to independently validate the calculated nonequilibrium free-energy by measuring it (or the performed work) in the course of the relaxation towards equilibrium?

We hope our work serves as a first step towards  the characterization of nonequilibrium thermodynamics in polymer processing. Its theoretical understanding should be useful to develop novel pathways  for targeted energy storage and release of polymers resulting from non-equilibrium stresses.~\cite{Reiter2020, Ramezani2020, Chandran2019} In other words, if the relaxation back to equilibrium can be controlled, the extra free energy stored in the system might be convertible into useful, time-dependent work. How this can be done in detail and how to define appropriately notions like heat and chemical work~\cite{Seifert2012} in such a process and what are the thermodynamic bounds and fluctuations relations~\cite{Evans1993, Jarzynski1997, Jarzynski1997b, Crooks1999} will be very interesting challenges for future studies. Furthermore, we hope that coarse-graining approaches for such nonequilibrium processes will help reducing the computational complexity of such systems,~\cite{Schilling2021} elucidating at the same time their intricate nonequilibrium thermodynamics.

\section*{Acknowledgements}
The authors thank Günter Reiter, J\"org Baschnagel, and Murugappan Muthukumar for inspiring discussions and useful
comments. The authors acknowledge support by the state of Baden-Württemberg through bwHPC
and the German Research Foundation (DFG) through grant no INST 39/963-1 FUGG (bw -
ForCluster NEMO). This work has been supported by the Deutsche Forschungsgemeinschaft (DFG) via grant WO 2410/2-1 within the framework of the Research Unit FOR 5099 ”Reducing complexity of nonequilibrium” (project No. 431945604). 

\section*{Data availability statement}
The data that support the findings of this study are available from the corresponding author upon reasonable request.

\section*{Supplementary Material}
See supplementary material for the derivation of $\Delta F_{\mathrm{eq}}$, details about the fitting approach for the polymer end-to-end distributions, and the extrapolation of density profiles of the free monomers around the active center.

\begin{thebibliography}{52}%
\makeatletter
\providecommand \@ifxundefined [1]{%
 \@ifx{#1\undefined}
}%
\providecommand \@ifnum [1]{%
 \ifnum #1\expandafter \@firstoftwo
 \else \expandafter \@secondoftwo
 \fi
}%
\providecommand \@ifx [1]{%
 \ifx #1\expandafter \@firstoftwo
 \else \expandafter \@secondoftwo
 \fi
}%
\providecommand \natexlab [1]{#1}%
\providecommand \enquote  [1]{``#1''}%
\providecommand \bibnamefont  [1]{#1}%
\providecommand \bibfnamefont [1]{#1}%
\providecommand \citenamefont [1]{#1}%
\providecommand \href@noop [0]{\@secondoftwo}%
\providecommand \href [0]{\begingroup \@sanitize@url \@href}%
\providecommand \@href[1]{\@@startlink{#1}\@@href}%
\providecommand \@@href[1]{\endgroup#1\@@endlink}%
\providecommand \@sanitize@url [0]{\catcode `\\12\catcode `\$12\catcode
  `\&12\catcode `\#12\catcode `\^12\catcode `\_12\catcode `\%12\relax}%
\providecommand \@@startlink[1]{}%
\providecommand \@@endlink[0]{}%
\providecommand \url  [0]{\begingroup\@sanitize@url \@url }%
\providecommand \@url [1]{\endgroup\@href {#1}{\urlprefix }}%
\providecommand \urlprefix  [0]{URL }%
\providecommand \Eprint [0]{\href }%
\providecommand \doibase [0]{http://dx.doi.org/}%
\providecommand \selectlanguage [0]{\@gobble}%
\providecommand \bibinfo  [0]{\@secondoftwo}%
\providecommand \bibfield  [0]{\@secondoftwo}%
\providecommand \translation [1]{[#1]}%
\providecommand \BibitemOpen [0]{}%
\providecommand \bibitemStop [0]{}%
\providecommand \bibitemNoStop [0]{.\EOS\space}%
\providecommand \EOS [0]{\spacefactor3000\relax}%
\providecommand \BibitemShut  [1]{\csname bibitem#1\endcsname}%
\let\auto@bib@innerbib\@empty
\bibitem [{\citenamefont {Bley}, \citenamefont {Baul},\ and\ \citenamefont
  {Dzubiella}(2021)}]{Bley2021}%
  \BibitemOpen
  \bibfield  {author} {\bibinfo {author} {\bibfnamefont {M.}~\bibnamefont
  {Bley}}, \bibinfo {author} {\bibfnamefont {U.}~\bibnamefont {Baul}}, \ and\
  \bibinfo {author} {\bibfnamefont {J.}~\bibnamefont {Dzubiella}},\ }\bibfield
  {title} {\enquote {\bibinfo {title} {{Controlling solvent quality by time:
  Self-avoiding sprints in nonequilibrium polymerization}},}\ }\href
  {https://journals.aps.org/pre/abstract/10.1103/PhysRevE.104.034501
  https://link.aps.org/doi/10.1103/PhysRevE.104.034501} {\bibfield  {journal}
  {\bibinfo  {journal} {Phys. Rev. E}\ }\textbf {\bibinfo {volume} {104}},\
  \bibinfo {pages} {034501} (\bibinfo {year} {2021})}\BibitemShut {NoStop}%
\bibitem [{\citenamefont {Rubinstein}\ and\ \citenamefont
  {Colby}(2003)}]{Rubinstein2003}%
  \BibitemOpen
  \bibfield  {author} {\bibinfo {author} {\bibfnamefont {M.}~\bibnamefont
  {Rubinstein}}\ and\ \bibinfo {author} {\bibfnamefont {R.~H.}\ \bibnamefont
  {Colby}},\ }\href@noop {} {\emph {\bibinfo {title} {{Polymer Physics}}}}\
  (\bibinfo  {publisher} {Oxford University Press},\ \bibinfo {address}
  {Oxford},\ \bibinfo {year} {2003})\BibitemShut {NoStop}%
\bibitem [{\citenamefont {Liphardt}(2002)}]{Liphardt2002}%
  \BibitemOpen
  \bibfield  {author} {\bibinfo {author} {\bibfnamefont {J.}~\bibnamefont
  {Liphardt}},\ }\bibfield  {title} {\enquote {\bibinfo {title} {{Equilibrium
  Information from Nonequilibrium Measurements in an Experimental Test of
  Jarzynski's Equality}},}\ }\href {\doibase 10.1126/science.1071152}
  {\bibfield  {journal} {\bibinfo  {journal} {Science}\ }\textbf {\bibinfo
  {volume} {296}},\ \bibinfo {pages} {1832--1835} (\bibinfo {year}
  {2002})}\BibitemShut {NoStop}%
\bibitem [{\citenamefont {Speck}\ and\ \citenamefont
  {Seifert}(2005)}]{Seifert2005}%
  \BibitemOpen
  \bibfield  {author} {\bibinfo {author} {\bibfnamefont {T.}~\bibnamefont
  {Speck}}\ and\ \bibinfo {author} {\bibfnamefont {U.}~\bibnamefont
  {Seifert}},\ }\bibfield  {title} {\enquote {\bibinfo {title} {{Dissipated
  work in driven harmonic diffusive systems: General solution and application
  to stretching Rouse polymers}},}\ }\href {\doibase
  10.1140/epjb/e2005-00086-6} {\bibfield  {journal} {\bibinfo  {journal} {Eur.
  Phys. J. B}\ }\textbf {\bibinfo {volume} {43}},\ \bibinfo {pages} {521--527}
  (\bibinfo {year} {2005})}\BibitemShut {NoStop}%
\bibitem [{\citenamefont {de~Gennes}(1979)}]{DeGennes1979}%
  \BibitemOpen
  \bibfield  {author} {\bibinfo {author} {\bibfnamefont {P.-G.}\ \bibnamefont
  {de~Gennes}},\ }\href@noop {} {\emph {\bibinfo {title} {{Scaling Concepts in
  Polymer Physics}}}}\ (\bibinfo  {publisher} {Cornell University Press},\
  \bibinfo {address} {Ithaca and London},\ \bibinfo {year} {1979})\BibitemShut
  {NoStop}%
\bibitem [{\citenamefont {Rouse}(1953)}]{Rouse1953}%
  \BibitemOpen
  \bibfield  {author} {\bibinfo {author} {\bibfnamefont {P.~E.}\ \bibnamefont
  {Rouse}},\ }\bibfield  {title} {\enquote {\bibinfo {title} {{A theory of the
  linear viscoelastic properties of dilute solutions of coiling polymers}},}\
  }\href {\doibase 10.1063/1.1699180} {\bibfield  {journal} {\bibinfo
  {journal} {J. Chem. Phys.}\ }\textbf {\bibinfo {volume} {21}},\ \bibinfo
  {pages} {1272--1280} (\bibinfo {year} {1953})}\BibitemShut {NoStop}%
\bibitem [{\citenamefont {Strobl}(2007)}]{Strobl2007}%
  \BibitemOpen
  \bibfield  {author} {\bibinfo {author} {\bibfnamefont {G.}~\bibnamefont
  {Strobl}},\ }\href@noop {} {\emph {\bibinfo {title} {{The physics of
  polymers}}}},\ \bibinfo {edition} {third rev.}\ ed.\ (\bibinfo  {publisher}
  {Springer Berlin Heidelberg},\ \bibinfo {address} {Berlin},\ \bibinfo {year}
  {2007})\BibitemShut {NoStop}%
\bibitem [{\citenamefont {Alexander-Katz}, \citenamefont {Wada},\ and\
  \citenamefont {Netz}(2009)}]{Alexander-Katz2009}%
  \BibitemOpen
  \bibfield  {author} {\bibinfo {author} {\bibfnamefont {A.}~\bibnamefont
  {Alexander-Katz}}, \bibinfo {author} {\bibfnamefont {H.}~\bibnamefont
  {Wada}}, \ and\ \bibinfo {author} {\bibfnamefont {R.~R.}\ \bibnamefont
  {Netz}},\ }\bibfield  {title} {\enquote {\bibinfo {title} {{Internal Friction
  and Nonequilibrium Unfolding of Polymeric Globules}},}\ }\href {\doibase
  10.1103/PhysRevLett.103.028102} {\bibfield  {journal} {\bibinfo  {journal}
  {Phys. Rev. Lett.}\ }\textbf {\bibinfo {volume} {103}},\ \bibinfo {pages}
  {028102} (\bibinfo {year} {2009})}\BibitemShut {NoStop}%
\bibitem [{\citenamefont {Einert}\ \emph {et~al.}(2011)\citenamefont {Einert},
  \citenamefont {Sing}, \citenamefont {Alexander-Katz},\ and\ \citenamefont
  {Netz}}]{Einert2011}%
  \BibitemOpen
  \bibfield  {author} {\bibinfo {author} {\bibfnamefont {T.~R.}\ \bibnamefont
  {Einert}}, \bibinfo {author} {\bibfnamefont {C.~E.}\ \bibnamefont {Sing}},
  \bibinfo {author} {\bibfnamefont {A.}~\bibnamefont {Alexander-Katz}}, \ and\
  \bibinfo {author} {\bibfnamefont {R.~R.}\ \bibnamefont {Netz}},\ }\bibfield
  {title} {\enquote {\bibinfo {title} {{Conformational dynamics and internal
  friction in homopolymer globules: equilibrium vs. non-equilibrium
  simulations}},}\ }\href {\doibase 10.1140/epje/i2011-11130-8} {\bibfield
  {journal} {\bibinfo  {journal} {Eur. Phys. J. E}\ }\textbf {\bibinfo {volume}
  {34}},\ \bibinfo {pages} {130} (\bibinfo {year} {2011})}\BibitemShut
  {NoStop}%
\bibitem [{\citenamefont {Latinwo}, \citenamefont {Hsiao},\ and\ \citenamefont
  {Schroeder}(2014)}]{Latinwo2014}%
  \BibitemOpen
  \bibfield  {author} {\bibinfo {author} {\bibfnamefont {F.}~\bibnamefont
  {Latinwo}}, \bibinfo {author} {\bibfnamefont {K.-W.}\ \bibnamefont {Hsiao}},
  \ and\ \bibinfo {author} {\bibfnamefont {C.~M.}\ \bibnamefont {Schroeder}},\
  }\bibfield  {title} {\enquote {\bibinfo {title} {{Nonequilibrium
  thermodynamics of dilute polymer solutions in flow}},}\ }\href {\doibase
  10.1063/1.4900880} {\bibfield  {journal} {\bibinfo  {journal} {J. Chem.
  Phys.}\ }\textbf {\bibinfo {volume} {141}},\ \bibinfo {pages} {174903}
  (\bibinfo {year} {2014})}\BibitemShut {NoStop}%
\bibitem [{\citenamefont {Reiter}(2020)}]{Reiter2020}%
  \BibitemOpen
  \bibfield  {author} {\bibinfo {author} {\bibfnamefont {G.}~\bibnamefont
  {Reiter}},\ }\bibfield  {title} {\enquote {\bibinfo {title} {{The memorizing
  capacity of polymers}},}\ }\href {\doibase 10.1063/1.5139621} {\bibfield
  {journal} {\bibinfo  {journal} {J. Chem. Phys.}\ }\textbf {\bibinfo {volume}
  {152}},\ \bibinfo {pages} {150901} (\bibinfo {year} {2020})}\BibitemShut
  {NoStop}%
\bibitem [{\citenamefont {Ramezani}, \citenamefont {Baschnagel},\ and\
  \citenamefont {Reiter}(2020)}]{Ramezani2020}%
  \BibitemOpen
  \bibfield  {author} {\bibinfo {author} {\bibfnamefont {F.}~\bibnamefont
  {Ramezani}}, \bibinfo {author} {\bibfnamefont {J.}~\bibnamefont
  {Baschnagel}}, \ and\ \bibinfo {author} {\bibfnamefont {G.}~\bibnamefont
  {Reiter}},\ }\bibfield  {title} {\enquote {\bibinfo {title} {{Translating
  molecular relaxations in non-equilibrated polymer melts into lifting
  macroscopic loads}},}\ }\href {\doibase 10.1103/PhysRevMaterials.4.082601}
  {\bibfield  {journal} {\bibinfo  {journal} {Phys. Rev. Mater.}\ }\textbf
  {\bibinfo {volume} {4}},\ \bibinfo {pages} {082601} (\bibinfo {year}
  {2020})}\BibitemShut {NoStop}%
\bibitem [{\citenamefont {Chandran}\ \emph {et~al.}(2019)\citenamefont
  {Chandran}, \citenamefont {Baschnagel}, \citenamefont {Cangialosi},
  \citenamefont {Fukao}, \citenamefont {Glynos}, \citenamefont {Janssen},
  \citenamefont {M{\"{u}}ller}, \citenamefont {Muthukumar}, \citenamefont
  {Steiner}, \citenamefont {Xu}, \citenamefont {Napolitano},\ and\
  \citenamefont {Reiter}}]{Chandran2019}%
  \BibitemOpen
  \bibfield  {author} {\bibinfo {author} {\bibfnamefont {S.}~\bibnamefont
  {Chandran}}, \bibinfo {author} {\bibfnamefont {J.}~\bibnamefont
  {Baschnagel}}, \bibinfo {author} {\bibfnamefont {D.}~\bibnamefont
  {Cangialosi}}, \bibinfo {author} {\bibfnamefont {K.}~\bibnamefont {Fukao}},
  \bibinfo {author} {\bibfnamefont {E.}~\bibnamefont {Glynos}}, \bibinfo
  {author} {\bibfnamefont {L.~M.~C.}\ \bibnamefont {Janssen}}, \bibinfo
  {author} {\bibfnamefont {M.}~\bibnamefont {M{\"{u}}ller}}, \bibinfo {author}
  {\bibfnamefont {M.}~\bibnamefont {Muthukumar}}, \bibinfo {author}
  {\bibfnamefont {U.}~\bibnamefont {Steiner}}, \bibinfo {author} {\bibfnamefont
  {J.}~\bibnamefont {Xu}}, \bibinfo {author} {\bibfnamefont {S.}~\bibnamefont
  {Napolitano}}, \ and\ \bibinfo {author} {\bibfnamefont {G.}~\bibnamefont
  {Reiter}},\ }\bibfield  {title} {\enquote {\bibinfo {title} {{Processing
  pathways decide polymer properties at the molecular level}},}\ }\href
  {\doibase 10.1021/acs.macromol.9b01195} {\bibfield  {journal} {\bibinfo
  {journal} {Macromolecules}\ }\textbf {\bibinfo {volume} {52}},\ \bibinfo
  {pages} {7146--7156} (\bibinfo {year} {2019})}\BibitemShut {NoStop}%
\bibitem [{\citenamefont {Stuart}\ \emph {et~al.}(2010)\citenamefont {Stuart},
  \citenamefont {Huck}, \citenamefont {Genzer}, \citenamefont {M{\"{u}}ller},
  \citenamefont {Ober}, \citenamefont {Stamm}, \citenamefont {Sukhorukov},
  \citenamefont {Szleifer}, \citenamefont {Tsukruk}, \citenamefont {Urban},
  \citenamefont {Winnik}, \citenamefont {Zauscher}, \citenamefont {Luzinov},\
  and\ \citenamefont {Minko}}]{Stuart2010}%
  \BibitemOpen
  \bibfield  {author} {\bibinfo {author} {\bibfnamefont {M.~A.~C.}\
  \bibnamefont {Stuart}}, \bibinfo {author} {\bibfnamefont {W.~T.~S.}\
  \bibnamefont {Huck}}, \bibinfo {author} {\bibfnamefont {J.}~\bibnamefont
  {Genzer}}, \bibinfo {author} {\bibfnamefont {M.}~\bibnamefont
  {M{\"{u}}ller}}, \bibinfo {author} {\bibfnamefont {C.}~\bibnamefont {Ober}},
  \bibinfo {author} {\bibfnamefont {M.}~\bibnamefont {Stamm}}, \bibinfo
  {author} {\bibfnamefont {G.~B.}\ \bibnamefont {Sukhorukov}}, \bibinfo
  {author} {\bibfnamefont {I.}~\bibnamefont {Szleifer}}, \bibinfo {author}
  {\bibfnamefont {V.~V.}\ \bibnamefont {Tsukruk}}, \bibinfo {author}
  {\bibfnamefont {M.}~\bibnamefont {Urban}}, \bibinfo {author} {\bibfnamefont
  {F.}~\bibnamefont {Winnik}}, \bibinfo {author} {\bibfnamefont
  {S.}~\bibnamefont {Zauscher}}, \bibinfo {author} {\bibfnamefont
  {I.}~\bibnamefont {Luzinov}}, \ and\ \bibinfo {author} {\bibfnamefont
  {S.}~\bibnamefont {Minko}},\ }\bibfield  {title} {\enquote {\bibinfo {title}
  {{Emerging applications of stimuli-responsive polymer materials}},}\ }\href
  {\doibase 10.1038/nmat2614} {\bibfield  {journal} {\bibinfo  {journal} {Nat.
  Mater.}\ }\textbf {\bibinfo {volume} {9}},\ \bibinfo {pages} {101--113}
  (\bibinfo {year} {2010})}\BibitemShut {NoStop}%
\bibitem [{\citenamefont {Thomas}\ \emph {et~al.}(2011)\citenamefont {Thomas},
  \citenamefont {Chenneviere}, \citenamefont {Reiter},\ and\ \citenamefont
  {Steiner}}]{Thomas2011}%
  \BibitemOpen
  \bibfield  {author} {\bibinfo {author} {\bibfnamefont {K.~R.}\ \bibnamefont
  {Thomas}}, \bibinfo {author} {\bibfnamefont {A.}~\bibnamefont {Chenneviere}},
  \bibinfo {author} {\bibfnamefont {G.}~\bibnamefont {Reiter}}, \ and\ \bibinfo
  {author} {\bibfnamefont {U.}~\bibnamefont {Steiner}},\ }\bibfield  {title}
  {\enquote {\bibinfo {title} {{Nonequilibrium behavior of thin polymer
  films}},}\ }\href {\doibase 10.1103/PhysRevE.83.021804} {\bibfield  {journal}
  {\bibinfo  {journal} {Phys. Rev. E}\ }\textbf {\bibinfo {volume} {83}},\
  \bibinfo {pages} {021804} (\bibinfo {year} {2011})}\BibitemShut {NoStop}%
\bibitem [{\citenamefont {Liu}\ \emph {et~al.}(2018)\citenamefont {Liu},
  \citenamefont {Yuk}, \citenamefont {Lin}, \citenamefont {Parada},
  \citenamefont {Tang}, \citenamefont {Tham}, \citenamefont {de~la
  Fuente-Nunez}, \citenamefont {Lu},\ and\ \citenamefont {Zhao}}]{Liu2018}%
  \BibitemOpen
  \bibfield  {author} {\bibinfo {author} {\bibfnamefont {X.}~\bibnamefont
  {Liu}}, \bibinfo {author} {\bibfnamefont {H.}~\bibnamefont {Yuk}}, \bibinfo
  {author} {\bibfnamefont {S.}~\bibnamefont {Lin}}, \bibinfo {author}
  {\bibfnamefont {G.~A.}\ \bibnamefont {Parada}}, \bibinfo {author}
  {\bibfnamefont {T.-C.}\ \bibnamefont {Tang}}, \bibinfo {author}
  {\bibfnamefont {E.}~\bibnamefont {Tham}}, \bibinfo {author} {\bibfnamefont
  {C.}~\bibnamefont {de~la Fuente-Nunez}}, \bibinfo {author} {\bibfnamefont
  {T.~K.}\ \bibnamefont {Lu}}, \ and\ \bibinfo {author} {\bibfnamefont
  {X.}~\bibnamefont {Zhao}},\ }\bibfield  {title} {\enquote {\bibinfo {title}
  {{3D printing of living responsive materials and devices}},}\ }\href
  {\doibase 10.1002/adma.201704821} {\bibfield  {journal} {\bibinfo  {journal}
  {Adv. Mater.}\ }\textbf {\bibinfo {volume} {30}},\ \bibinfo {pages} {1704821}
  (\bibinfo {year} {2018})}\BibitemShut {NoStop}%
\bibitem [{\citenamefont {Walther}(2020)}]{Walther2019}%
  \BibitemOpen
  \bibfield  {author} {\bibinfo {author} {\bibfnamefont {A.}~\bibnamefont
  {Walther}},\ }\bibfield  {title} {\enquote {\bibinfo {title} {{Viewpoint:
  From responsive to adaptive and interactive materials and materials systems:
  A roadmap}},}\ }\href {\doibase 10.1002/adma.201905111} {\bibfield  {journal}
  {\bibinfo  {journal} {Adv. Mater.}\ }\textbf {\bibinfo {volume} {32}},\
  \bibinfo {pages} {1905111} (\bibinfo {year} {2020})}\BibitemShut {NoStop}%
\bibitem [{\citenamefont {de~Gennes}(1982)}]{DeGennes1982a}%
  \BibitemOpen
  \bibfield  {author} {\bibinfo {author} {\bibfnamefont {P.-G.}\ \bibnamefont
  {de~Gennes}},\ }\bibfield  {title} {\enquote {\bibinfo {title} {{Kinetics of
  diffusion‐controlled processes in dense polymer systems. I. Nonentangled
  regimes}},}\ }\href {\doibase 10.1063/1.443328} {\bibfield  {journal}
  {\bibinfo  {journal} {J. Chem. Phys.}\ }\textbf {\bibinfo {volume} {76}},\
  \bibinfo {pages} {3316--3321} (\bibinfo {year} {1982})}\BibitemShut {NoStop}%
\bibitem [{\citenamefont {Gu{\'{e}}rin}, \citenamefont {B{\'{e}}nichou},\ and\
  \citenamefont {Voituriez}(2012)}]{Guerin2012}%
  \BibitemOpen
  \bibfield  {author} {\bibinfo {author} {\bibfnamefont {T.}~\bibnamefont
  {Gu{\'{e}}rin}}, \bibinfo {author} {\bibfnamefont {O.}~\bibnamefont
  {B{\'{e}}nichou}}, \ and\ \bibinfo {author} {\bibfnamefont {R.}~\bibnamefont
  {Voituriez}},\ }\bibfield  {title} {\enquote {\bibinfo {title}
  {{Non-Markovian polymer reaction kinetics}},}\ }\href {\doibase
  10.1038/nchem.1378} {\bibfield  {journal} {\bibinfo  {journal} {Nat. Chem.}\
  }\textbf {\bibinfo {volume} {4}},\ \bibinfo {pages} {568--573} (\bibinfo
  {year} {2012})}\BibitemShut {NoStop}%
\bibitem [{\citenamefont {Katkar}\ and\ \citenamefont
  {Muthukumar}(2018)}]{Katkar2018}%
  \BibitemOpen
  \bibfield  {author} {\bibinfo {author} {\bibfnamefont {H.~H.}\ \bibnamefont
  {Katkar}}\ and\ \bibinfo {author} {\bibfnamefont {M.}~\bibnamefont
  {Muthukumar}},\ }\bibfield  {title} {\enquote {\bibinfo {title} {{Role of
  non-equilibrium conformations on driven polymer translocation}},}\ }\href
  {\doibase 10.1063/1.4994204} {\bibfield  {journal} {\bibinfo  {journal} {J.
  Chem. Phys.}\ }\textbf {\bibinfo {volume} {148}},\ \bibinfo {pages} {024903}
  (\bibinfo {year} {2018})}\BibitemShut {NoStop}%
\bibitem [{\citenamefont {Chubak}\ \emph {et~al.}(2020)\citenamefont {Chubak},
  \citenamefont {Likos}, \citenamefont {Kremer},\ and\ \citenamefont
  {Smrek}}]{Chubak2020}%
  \BibitemOpen
  \bibfield  {author} {\bibinfo {author} {\bibfnamefont {I.}~\bibnamefont
  {Chubak}}, \bibinfo {author} {\bibfnamefont {C.~N.}\ \bibnamefont {Likos}},
  \bibinfo {author} {\bibfnamefont {K.}~\bibnamefont {Kremer}}, \ and\ \bibinfo
  {author} {\bibfnamefont {J.}~\bibnamefont {Smrek}},\ }\bibfield  {title}
  {\enquote {\bibinfo {title} {{Emergence of active topological glass through
  directed chain dynamics and nonequilibrium phase segregation}},}\ }\href
  {\doibase 10.1103/PhysRevResearch.2.043249} {\bibfield  {journal} {\bibinfo
  {journal} {Phys. Rev. Res.}\ }\textbf {\bibinfo {volume} {2}},\ \bibinfo
  {pages} {043249} (\bibinfo {year} {2020})}\BibitemShut {NoStop}%
\bibitem [{\citenamefont {{De Groot}}\ and\ \citenamefont
  {Mazur}(1962)}]{DeGroot1962}%
  \BibitemOpen
  \bibfield  {author} {\bibinfo {author} {\bibfnamefont {S.~R.}\ \bibnamefont
  {{De Groot}}}\ and\ \bibinfo {author} {\bibfnamefont {P.}~\bibnamefont
  {Mazur}},\ }\href@noop {} {\emph {\bibinfo {title} {{Non-equilibrium
  thermodynamics}}}}\ (\bibinfo  {publisher} {Dover Publications},\ \bibinfo
  {address} {New York},\ \bibinfo {year} {1962})\BibitemShut {NoStop}%
\bibitem [{\citenamefont {Prigogine}(1978)}]{Prigogine1978}%
  \BibitemOpen
  \bibfield  {author} {\bibinfo {author} {\bibfnamefont {I.}~\bibnamefont
  {Prigogine}},\ }\bibfield  {title} {\enquote {\bibinfo {title} {{Time,
  structure, and fluctuations}},}\ }\href {\doibase
  10.1126/science.201.4358.777} {\bibfield  {journal} {\bibinfo  {journal}
  {Science}\ }\textbf {\bibinfo {volume} {201}},\ \bibinfo {pages} {777--785}
  (\bibinfo {year} {1978})}\BibitemShut {NoStop}%
\bibitem [{\citenamefont {Seifert}(2005)}]{Seifert2005b}%
  \BibitemOpen
  \bibfield  {author} {\bibinfo {author} {\bibfnamefont {U.}~\bibnamefont
  {Seifert}},\ }\bibfield  {title} {\enquote {\bibinfo {title} {{Entropy
  production along a stochastic trajectory and an integral fluctuation
  theorem}},}\ }\href {\doibase 10.1103/PhysRevLett.95.040602} {\bibfield
  {journal} {\bibinfo  {journal} {Phys. Rev. Lett.}\ }\textbf {\bibinfo
  {volume} {95}},\ \bibinfo {pages} {040602} (\bibinfo {year}
  {2005})}\BibitemShut {NoStop}%
\bibitem [{\citenamefont {Seifert}(2012)}]{Seifert2012}%
  \BibitemOpen
  \bibfield  {author} {\bibinfo {author} {\bibfnamefont {U.}~\bibnamefont
  {Seifert}},\ }\bibfield  {title} {\enquote {\bibinfo {title} {{Stochastic
  thermodynamics, fluctuation theorems and molecular machines}},}\ }\href
  {\doibase 10.1088/0034-4885/75/12/126001} {\bibfield  {journal} {\bibinfo
  {journal} {Reports Prog. Phys.}\ }\textbf {\bibinfo {volume} {75}},\ \bibinfo
  {pages} {126001} (\bibinfo {year} {2012})}\BibitemShut {NoStop}%
\bibitem [{\citenamefont {Evans}, \citenamefont {Cohen},\ and\ \citenamefont
  {Morriss}(1993)}]{Evans1993}%
  \BibitemOpen
  \bibfield  {author} {\bibinfo {author} {\bibfnamefont {D.~J.}\ \bibnamefont
  {Evans}}, \bibinfo {author} {\bibfnamefont {E.~G.~D.}\ \bibnamefont {Cohen}},
  \ and\ \bibinfo {author} {\bibfnamefont {G.~P.}\ \bibnamefont {Morriss}},\
  }\bibfield  {title} {\enquote {\bibinfo {title} {{Probability of second law
  violations in shearing steady states}},}\ }\href {\doibase
  10.1103/PhysRevLett.71.2401} {\bibfield  {journal} {\bibinfo  {journal}
  {Phys. Rev. Lett.}\ }\textbf {\bibinfo {volume} {71}},\ \bibinfo {pages}
  {2401--2404} (\bibinfo {year} {1993})}\BibitemShut {NoStop}%
\bibitem [{\citenamefont {Jarzynski}(1997{\natexlab{a}})}]{Jarzynski1997}%
  \BibitemOpen
  \bibfield  {author} {\bibinfo {author} {\bibfnamefont {C.}~\bibnamefont
  {Jarzynski}},\ }\bibfield  {title} {\enquote {\bibinfo {title} {{Equilibrium
  free-energy differences from nonequilibrium measurements: A master-equation
  approach}},}\ }\href {\doibase 10.1103/PhysRevE.56.5018} {\bibfield
  {journal} {\bibinfo  {journal} {Phys. Rev. E}\ }\textbf {\bibinfo {volume}
  {56}},\ \bibinfo {pages} {5018--5035} (\bibinfo {year}
  {1997}{\natexlab{a}})}\BibitemShut {NoStop}%
\bibitem [{\citenamefont {Jarzynski}(1997{\natexlab{b}})}]{Jarzynski1997b}%
  \BibitemOpen
  \bibfield  {author} {\bibinfo {author} {\bibfnamefont {C.}~\bibnamefont
  {Jarzynski}},\ }\bibfield  {title} {\enquote {\bibinfo {title}
  {{Nonequilibrium equality for free energy differences}},}\ }\href {\doibase
  10.1103/PhysRevLett.78.2690} {\bibfield  {journal} {\bibinfo  {journal}
  {Phys. Rev. Lett.}\ }\textbf {\bibinfo {volume} {78}},\ \bibinfo {pages}
  {2690--2693} (\bibinfo {year} {1997}{\natexlab{b}})}\BibitemShut {NoStop}%
\bibitem [{\citenamefont {Crooks}(1999)}]{Crooks1999}%
  \BibitemOpen
  \bibfield  {author} {\bibinfo {author} {\bibfnamefont {G.~E.}\ \bibnamefont
  {Crooks}},\ }\bibfield  {title} {\enquote {\bibinfo {title} {{Entropy
  production fluctuation theorem and the nonequilibrium work relation for free
  energy differences}},}\ }\href {\doibase 10.1103/PhysRevE.60.2721} {\bibfield
   {journal} {\bibinfo  {journal} {Phys. Rev. E}\ }\textbf {\bibinfo {volume}
  {60}},\ \bibinfo {pages} {2721--2726} (\bibinfo {year} {1999})}\BibitemShut
  {NoStop}%
\bibitem [{\citenamefont {Gibbs}(1902)}]{Gibbs1902}%
  \BibitemOpen
  \bibfield  {author} {\bibinfo {author} {\bibfnamefont {J.~W.}\ \bibnamefont
  {Gibbs}},\ }\href@noop {} {\emph {\bibinfo {title} {{Elementary Principles in
  Statistical Mechanics}}}}\ (\bibinfo  {publisher} {Charles Scribner's Sons},\
  \bibinfo {address} {New York, New York, USA},\ \bibinfo {year}
  {1902})\BibitemShut {NoStop}%
\bibitem [{\citenamefont {Shannon}(1948)}]{Shannon1948}%
  \BibitemOpen
  \bibfield  {author} {\bibinfo {author} {\bibfnamefont {C.~E.}\ \bibnamefont
  {Shannon}},\ }\bibfield  {title} {\enquote {\bibinfo {title} {{A mathematical
  theory of communication}},}\ }\href {\doibase
  10.1002/j.1538-7305.1948.tb00917.x} {\bibfield  {journal} {\bibinfo
  {journal} {Bell Syst. Tech. J.}\ }\textbf {\bibinfo {volume} {27}},\ \bibinfo
  {pages} {623--656} (\bibinfo {year} {1948})}\BibitemShut {NoStop}%
\bibitem [{\citenamefont {Kullback}\ and\ \citenamefont
  {Leibler}(1951)}]{Kullback1951}%
  \BibitemOpen
  \bibfield  {author} {\bibinfo {author} {\bibfnamefont {S.}~\bibnamefont
  {Kullback}}\ and\ \bibinfo {author} {\bibfnamefont {R.~A.}\ \bibnamefont
  {Leibler}},\ }\bibfield  {title} {\enquote {\bibinfo {title} {{On information
  and sufficiency}},}\ }\href {\doibase 10.1214/aoms/1177729694} {\bibfield
  {journal} {\bibinfo  {journal} {Ann. Math. Stat.}\ }\textbf {\bibinfo
  {volume} {22}},\ \bibinfo {pages} {79--86} (\bibinfo {year}
  {1951})}\BibitemShut {NoStop}%
\bibitem [{\citenamefont {Altaner}(2017)}]{Altaner2017}%
  \BibitemOpen
  \bibfield  {author} {\bibinfo {author} {\bibfnamefont {B.}~\bibnamefont
  {Altaner}},\ }\bibfield  {title} {\enquote {\bibinfo {title} {{Nonequilibrium
  thermodynamics and information theory: basic concepts and relaxing
  dynamics}},}\ }\href {\doibase 10.1088/1751-8121/aa841d} {\bibfield
  {journal} {\bibinfo  {journal} {J. Phys. A Math. Theor.}\ }\textbf {\bibinfo
  {volume} {50}},\ \bibinfo {pages} {454001} (\bibinfo {year}
  {2017})}\BibitemShut {NoStop}%
\bibitem [{\citenamefont {Wall}(1942)}]{Wall1942}%
  \BibitemOpen
  \bibfield  {author} {\bibinfo {author} {\bibfnamefont {F.~T.}\ \bibnamefont
  {Wall}},\ }\bibfield  {title} {\enquote {\bibinfo {title} {{Statistical
  thermodynamics of rubber}},}\ }\href {\doibase 10.1063/1.1723668} {\bibfield
  {journal} {\bibinfo  {journal} {J. Chem. Phys.}\ }\textbf {\bibinfo {volume}
  {10}},\ \bibinfo {pages} {132--134} (\bibinfo {year} {1942})}\BibitemShut
  {NoStop}%
\bibitem [{\citenamefont {Dayantis}(1995)}]{Dayantis1994}%
  \BibitemOpen
  \bibfield  {author} {\bibinfo {author} {\bibfnamefont {J.}~\bibnamefont
  {Dayantis}},\ }\bibfield  {title} {\enquote {\bibinfo {title} {{On the
  entropy of single polymer chains}},}\ }\href {\doibase
  10.1016/0014-3057(94)00143-X} {\bibfield  {journal} {\bibinfo  {journal}
  {Eur. Polym. J.}\ }\textbf {\bibinfo {volume} {31}},\ \bibinfo {pages}
  {203--204} (\bibinfo {year} {1995})}\BibitemShut {NoStop}%
\bibitem [{\citenamefont {Qian}(2001)}]{Qian2001}%
  \BibitemOpen
  \bibfield  {author} {\bibinfo {author} {\bibfnamefont {H.}~\bibnamefont
  {Qian}},\ }\bibfield  {title} {\enquote {\bibinfo {title} {{Relative entropy:
  Free energy associated with equilibrium fluctuations and nonequilibrium
  deviations}},}\ }\href {\doibase 10.1103/PhysRevE.63.042103} {\bibfield
  {journal} {\bibinfo  {journal} {Phys. Rev. E - Stat. Physics, Plasmas,
  Fluids, Relat. Interdiscip. Top.}\ }\textbf {\bibinfo {volume} {63}},\
  \bibinfo {pages} {1--4} (\bibinfo {year} {2001})}\BibitemShut {NoStop}%
\bibitem [{\citenamefont {Edwards}, \citenamefont {{Nafar Sefiddashti}},\ and\
  \citenamefont {Khomami}(2021)}]{Edwards2021}%
  \BibitemOpen
  \bibfield  {author} {\bibinfo {author} {\bibfnamefont {B.~J.}\ \bibnamefont
  {Edwards}}, \bibinfo {author} {\bibfnamefont {M.~H.}\ \bibnamefont {{Nafar
  Sefiddashti}}}, \ and\ \bibinfo {author} {\bibfnamefont {B.}~\bibnamefont
  {Khomami}},\ }\bibfield  {title} {\enquote {\bibinfo {title} {{A method for
  calculating the nonequilibrium entropy of a flowing polymer melt via
  atomistic simulation}},}\ }\href {\doibase 10.1063/5.0056547} {\bibfield
  {journal} {\bibinfo  {journal} {J. Chem. Phys.}\ }\textbf {\bibinfo {volume}
  {155}},\ \bibinfo {pages} {111101} (\bibinfo {year} {2021})}\BibitemShut
  {NoStop}%
\bibitem [{\citenamefont {Lahiri}\ \emph {et~al.}(2015)\citenamefont {Lahiri},
  \citenamefont {Wang}, \citenamefont {Esposito},\ and\ \citenamefont
  {Lacoste}}]{Lahiri2015}%
  \BibitemOpen
  \bibfield  {author} {\bibinfo {author} {\bibfnamefont {S.}~\bibnamefont
  {Lahiri}}, \bibinfo {author} {\bibfnamefont {Y.}~\bibnamefont {Wang}},
  \bibinfo {author} {\bibfnamefont {M.}~\bibnamefont {Esposito}}, \ and\
  \bibinfo {author} {\bibfnamefont {D.}~\bibnamefont {Lacoste}},\ }\bibfield
  {title} {\enquote {\bibinfo {title} {{Kinetics and thermodynamics of
  reversible polymerization in closed systems}},}\ }\href {\doibase
  10.1088/1367-2630/17/8/085008} {\bibfield  {journal} {\bibinfo  {journal}
  {New J. Phys.}\ }\textbf {\bibinfo {volume} {17}},\ \bibinfo {pages} {085008}
  (\bibinfo {year} {2015})}\BibitemShut {NoStop}%
\bibitem [{\citenamefont {Gavrilov}, \citenamefont {Ch{\'{e}}trite},\ and\
  \citenamefont {Bechhoefer}(2017)}]{Gavrilov2017}%
  \BibitemOpen
  \bibfield  {author} {\bibinfo {author} {\bibfnamefont {M.}~\bibnamefont
  {Gavrilov}}, \bibinfo {author} {\bibfnamefont {R.}~\bibnamefont
  {Ch{\'{e}}trite}}, \ and\ \bibinfo {author} {\bibfnamefont {J.}~\bibnamefont
  {Bechhoefer}},\ }\bibfield  {title} {\enquote {\bibinfo {title} {{Direct
  measurement of weakly nonequilibrium system entropy is consistent with
  Gibbs–Shannon form}},}\ }\href {\doibase 10.1073/pnas.1708689114}
  {\bibfield  {journal} {\bibinfo  {journal} {Proc. Natl. Acad. Sci.}\ }\textbf
  {\bibinfo {volume} {114}},\ \bibinfo {pages} {11097--11102} (\bibinfo {year}
  {2017})}\BibitemShut {NoStop}%
\bibitem [{\citenamefont {Plimpton}(1995)}]{Plimpton1995}%
  \BibitemOpen
  \bibfield  {author} {\bibinfo {author} {\bibfnamefont {S.}~\bibnamefont
  {Plimpton}},\ }\bibfield  {title} {\enquote {\bibinfo {title} {{Fast parallel
  algorithms for short-range molecular dynamics}},}\ }\href {\doibase
  10.1006/jcph.1995.1039} {\bibfield  {journal} {\bibinfo  {journal} {J.
  Comput. Phys.}\ }\textbf {\bibinfo {volume} {117}},\ \bibinfo {pages} {1--19}
  (\bibinfo {year} {1995})}\BibitemShut {NoStop}%
\bibitem [{\citenamefont {Doi}\ and\ \citenamefont {Edwards}(1986)}]{Doi1986}%
  \BibitemOpen
  \bibfield  {author} {\bibinfo {author} {\bibfnamefont {M.}~\bibnamefont
  {Doi}}\ and\ \bibinfo {author} {\bibfnamefont {S.~F.}\ \bibnamefont
  {Edwards}},\ }\href@noop {} {\emph {\bibinfo {title} {{The Theory of Polymer
  Dynamics}}}}\ (\bibinfo  {publisher} {Oxford University Press},\ \bibinfo
  {address} {Oxford},\ \bibinfo {year} {1986})\BibitemShut {NoStop}%
\bibitem [{\citenamefont {Doi}(1975)}]{Doi1975a}%
  \BibitemOpen
  \bibfield  {author} {\bibinfo {author} {\bibfnamefont {M.}~\bibnamefont
  {Doi}},\ }\bibfield  {title} {\enquote {\bibinfo {title} {{Theory of
  diffusion-controlled reaction between non-simple molecules. I}},}\ }\href
  {\doibase 10.1016/0301-0104(75)80043-7} {\bibfield  {journal} {\bibinfo
  {journal} {Chem. Phys.}\ }\textbf {\bibinfo {volume} {11}},\ \bibinfo {pages}
  {107--113} (\bibinfo {year} {1975})}\BibitemShut {NoStop}%
\bibitem [{\citenamefont {Erban}\ and\ \citenamefont
  {Chapman}(2009)}]{Erban2009}%
  \BibitemOpen
  \bibfield  {author} {\bibinfo {author} {\bibfnamefont {R.}~\bibnamefont
  {Erban}}\ and\ \bibinfo {author} {\bibfnamefont {S.~J.}\ \bibnamefont
  {Chapman}},\ }\bibfield  {title} {\enquote {\bibinfo {title} {{Stochastic
  modelling of reaction–diffusion processes: algorithms for bimolecular
  reactions}},}\ }\href {\doibase 10.1088/1478-3975/6/4/046001} {\bibfield
  {journal} {\bibinfo  {journal} {Phys. Biol.}\ }\textbf {\bibinfo {volume}
  {6}},\ \bibinfo {pages} {046001} (\bibinfo {year} {2009})}\BibitemShut
  {NoStop}%
\bibitem [{\citenamefont {de~Buyl}\ and\ \citenamefont
  {Nies}(2015)}]{DeBuyl2015}%
  \BibitemOpen
  \bibfield  {author} {\bibinfo {author} {\bibfnamefont {P.}~\bibnamefont
  {de~Buyl}}\ and\ \bibinfo {author} {\bibfnamefont {E.}~\bibnamefont {Nies}},\
  }\bibfield  {title} {\enquote {\bibinfo {title} {{A parallel algorithm for
  step- and chain-growth polymerization in molecular dynamics}},}\ }\href
  {\doibase 10.1063/1.4916313} {\bibfield  {journal} {\bibinfo  {journal} {J.
  Chem. Phys.}\ }\textbf {\bibinfo {volume} {142}},\ \bibinfo {pages} {134102}
  (\bibinfo {year} {2015})}\BibitemShut {NoStop}%
\bibitem [{\citenamefont {Glatting}, \citenamefont {Winkler},\ and\
  \citenamefont {Reineker}(1993)}]{Winkler}%
  \BibitemOpen
  \bibfield  {author} {\bibinfo {author} {\bibfnamefont {G.}~\bibnamefont
  {Glatting}}, \bibinfo {author} {\bibfnamefont {R.~G.}\ \bibnamefont
  {Winkler}}, \ and\ \bibinfo {author} {\bibfnamefont {P.}~\bibnamefont
  {Reineker}},\ }\bibfield  {title} {\enquote {\bibinfo {title} {Partition
  function and force extension relation for a generalized freely jointed
  chain},}\ }\href@noop {} {\bibfield  {journal} {\bibinfo  {journal}
  {Macromolecules}\ }\textbf {\bibinfo {volume} {26}},\ \bibinfo {pages}
  {6085--6091} (\bibinfo {year} {1993})}\BibitemShut {NoStop}%
\bibitem [{\citenamefont {Manca}\ \emph {et~al.}(2012)\citenamefont {Manca},
  \citenamefont {Giordano}, \citenamefont {Palla}, \citenamefont {Zucca},
  \citenamefont {Cleri},\ and\ \citenamefont {Colombo}}]{Colombo}%
  \BibitemOpen
  \bibfield  {author} {\bibinfo {author} {\bibfnamefont {F.}~\bibnamefont
  {Manca}}, \bibinfo {author} {\bibfnamefont {S.}~\bibnamefont {Giordano}},
  \bibinfo {author} {\bibfnamefont {P.~L.}\ \bibnamefont {Palla}}, \bibinfo
  {author} {\bibfnamefont {R.}~\bibnamefont {Zucca}}, \bibinfo {author}
  {\bibfnamefont {F.}~\bibnamefont {Cleri}}, \ and\ \bibinfo {author}
  {\bibfnamefont {L.}~\bibnamefont {Colombo}},\ }\bibfield  {title} {\enquote
  {\bibinfo {title} {Elasticity of flexible and semiflexible polymers with
  extensible bonds in the {G}ibbs and {H}elmholtz ensembles},}\ }\href@noop {}
  {\bibfield  {journal} {\bibinfo  {journal} {J. Chem. Phys.}\ }\textbf
  {\bibinfo {volume} {136}},\ \bibinfo {pages} {154906} (\bibinfo {year}
  {2012})}\BibitemShut {NoStop}%
\bibitem [{\citenamefont {Fiasconaro}\ and\ \citenamefont
  {Falo}(2019)}]{Fiasco}%
  \BibitemOpen
  \bibfield  {author} {\bibinfo {author} {\bibfnamefont {A.}~\bibnamefont
  {Fiasconaro}}\ and\ \bibinfo {author} {\bibfnamefont {F.}~\bibnamefont
  {Falo}},\ }\bibfield  {title} {\enquote {\bibinfo {title} {Analytical results
  of the extensible freely jointed chain model},}\ }\href@noop {} {\bibfield
  {journal} {\bibinfo  {journal} {Physica A}\ }\textbf {\bibinfo {volume}
  {532}},\ \bibinfo {pages} {121929} (\bibinfo {year} {2019})}\BibitemShut
  {NoStop}%
\bibitem [{\citenamefont {McQuarrie}(1976)}]{McQuarrie1976}%
  \BibitemOpen
  \bibfield  {author} {\bibinfo {author} {\bibfnamefont {D.~A.}\ \bibnamefont
  {McQuarrie}},\ }\href@noop {} {\emph {\bibinfo {title} {{Statistical
  mechanics}}}}\ (\bibinfo  {publisher} {Harper {\&} Row},\ \bibinfo {address}
  {New York},\ \bibinfo {year} {1976})\BibitemShut {NoStop}%
\bibitem [{\citenamefont {Atkins}\ and\ \citenamefont
  {de~Paula}(2010)}]{Atkins2010}%
  \BibitemOpen
  \bibfield  {author} {\bibinfo {author} {\bibfnamefont {P.~W.}\ \bibnamefont
  {Atkins}}\ and\ \bibinfo {author} {\bibfnamefont {J.}~\bibnamefont
  {de~Paula}},\ }\href@noop {} {\emph {\bibinfo {title} {{Physical
  Chemistry}}}},\ \bibinfo {edition} {9th}\ ed.\ (\bibinfo  {publisher} {W.H.
  Freeman and Company},\ \bibinfo {address} {New York},\ \bibinfo {year}
  {2010})\BibitemShut {NoStop}%
\bibitem [{\citenamefont {von Smoluchowski}(1918)}]{Smoluchowski1918}%
  \BibitemOpen
  \bibfield  {author} {\bibinfo {author} {\bibfnamefont {M.}~\bibnamefont {von
  Smoluchowski}},\ }\bibfield  {title} {\enquote {\bibinfo {title} {{Versuch
  einer mathematischen Theorie der Koagulationskinetik kolloider
  L{\"{o}}sungen}},}\ }\href {\doibase 10.1515/zpch-1918-9209} {\bibfield
  {journal} {\bibinfo  {journal} {Zeitschrift f{\"{u}}r Phys. Chemie}\ }\textbf
  {\bibinfo {volume} {92U}},\ \bibinfo {pages} {129--168} (\bibinfo {year}
  {1918})}\BibitemShut {NoStop}%
\bibitem [{\citenamefont {Dibak}\ \emph {et~al.}(2019)\citenamefont {Dibak},
  \citenamefont {Fr{\"{o}}hner}, \citenamefont {No{\'{e}}},\ and\ \citenamefont
  {H{\"{o}}fling}}]{Dibak2019}%
  \BibitemOpen
  \bibfield  {author} {\bibinfo {author} {\bibfnamefont {M.}~\bibnamefont
  {Dibak}}, \bibinfo {author} {\bibfnamefont {C.}~\bibnamefont
  {Fr{\"{o}}hner}}, \bibinfo {author} {\bibfnamefont {F.}~\bibnamefont
  {No{\'{e}}}}, \ and\ \bibinfo {author} {\bibfnamefont {F.}~\bibnamefont
  {H{\"{o}}fling}},\ }\bibfield  {title} {\enquote {\bibinfo {title}
  {{Diffusion-influenced reaction rates in the presence of pair
  interactions}},}\ }\href {\doibase 10.1063/1.5124728} {\bibfield  {journal}
  {\bibinfo  {journal} {J. Chem. Phys.}\ }\textbf {\bibinfo {volume} {151}},\
  \bibinfo {pages} {164105} (\bibinfo {year} {2019})}\BibitemShut {NoStop}%
\bibitem [{\citenamefont {Schilling}(2021)}]{Schilling2021}%
  \BibitemOpen
  \bibfield  {author} {\bibinfo {author} {\bibfnamefont {T.}~\bibnamefont
  {Schilling}},\ }\bibfield  {title} {\enquote {\bibinfo {title}
  {{Coarse-Grained Modelling Out of Equilibrium}},}\ }\href
  {http://arxiv.org/abs/2107.09972} {\  (\bibinfo {year} {2021})},\ \Eprint
  {http://arxiv.org/abs/2107.09972} {arXiv:2107.09972} \BibitemShut {NoStop}%
\end{thebibliography}
%

\end{document}